\documentclass[
showpacs,floatfix, aps,prb,amsmath,
twocolumn,
groupaddress,eqsecnum]{revtex4}

\usepackage{epsfig}
\usepackage{epsf}
\usepackage{array}
\newcommand{\beq}{\begin{equation}}
\newcommand{\eeq}{\end{equation}}
\newcommand{\be}{\begin{equation}}
\newcommand{\ee}{\end{equation}}
\newcommand{\beqa}{\begin{eqnarray}}
\newcommand{\eeqa}{\end{eqnarray}}
\newcommand{\bea}{\begin{eqnarray}}
\newcommand{\eea}{\end{eqnarray}}
\newcommand{\w}{\omega}

\newcommand{\nn}{\mbox{\boldmath $i$}}
\newcommand{\mm}{\mbox{\boldmath $e$}}
\renewcommand{\r}{\mbox{\boldmath $r$}}
\renewcommand{\P}{\mbox{\boldmath $P$}}
\newcommand{\p}{\mbox{\boldmath $p$}}
\newcommand{\R}{\mbox{\boldmath $R$}}

\renewcommand{\v}{\mbox{\boldmath $v$}}
\newcommand{\E}{\mbox{\boldmath $E$}}
\renewcommand{\j}{\mbox{\boldmath $j$}}
\newcommand{\nnabla}{\mbox{\boldmath $\nabla$}}

\newcommand{\q}{\mbox{\boldmath $q$}}

\newcommand{\pt}{\partial_t}

\newcommand{\bzeta}{\mbox{\boldmath $\zeta$}}

\newcommand{\St}{\mbox{St}}

\renewcommand{\Re}{\mbox{$\mathrm Re$} }

\newcommand{\mylabel}[1]{\label{#1}
}
\newcommand{\req}[1]{Eq.~(\ref{#1})}
\newcommand{\reqs}[1]{Eqs.~(\ref{#1})}
\newcommand{\rref}[1]{(\ref{#1})}
\newcommand{\ep}{\varepsilon}
\newcommand{\lH}{\lambda_H}

\begin{document}

\title{Magnetotransport in  two-dimensional electron gas
at large filling factors}

\author{M.G. Vavilov}
\affiliation{Theoretical Physics Institute, University of
Minnesota, Minneapolis, MN 55455 }

\author{I.L.
  Aleiner}
 \affiliation{ Physics Department, Columbia University, New
  York, NY 10027 }

\pacs{73.40.-c, 73.50.Pz, 73.43.Qt, 73.50.Fq }
\begin{abstract}
We derive the quantum Boltzmann equation  for the two-dimensional
electron gas in a magnetic field such that the filling factor $\nu
\gg 1$. This equation describes all  of the effects of the
external fields on the impurity collision integral including
Shubnikov-de Haas oscillations, smooth part of the
magnetoresistance, and non-linear transport. Furthermore,
we obtain quantitative
results for the effect of the external microwave radiation on the
linear and non-linear $dc$ transport in the system. Our findings are
relevant for the description of the oscillating resistivity
discovered by Zudov {\em et al.}, zero-resistance state
discovered by  Mani {\em et al.} and Zudov {\em et al.}, and for
the microscopic justification of the model of Andreev {\em et
al.}. We also present semiclassical picture for the qualitative
consideration of the effects of the applied field on the collision
integral.
\end{abstract}
\date{\today}
\maketitle

\section{Introduction}
\label{intro}

The purpose of this paper is to construct the
theory describing linear magnetotransport, non-linear effect of $dc$
electric field, and effect of microwave on both linear and nonlinear $dc$
magnetotransport from a unified point of view.
Despite a long history of the systematic
experimental and
theoretical study of the properties of two-dimensional electron
and hole systems in
semiclassically strong\cite{Andoreview} and quantizing magnetic
field\cite{QHEreview}, the system still brings us  surprises.

Recent experiments
\cite{Zudov1,Mani,Zudov2,Mani2,Dorozhkin,Zudov4} revealed the
new class  of phenomena. (In fact, such effects were first
considered theoretically by Ryzhii\cite{Ryzhii,Ryzhii1}
three decades ago but were
not fully appreciated.)
Exposing the two-dimensional
electron system to a microwave radiation, Zudov {\em et al.}\cite{Zudov1}
discovered the drastic oscillations of the longitudinal resistivity as
a function of the magnetic field. The period of these oscillations
was controlled only by the ratio of the microwave frequency $\omega $ to
the cyclotron frequency $\omega _{c}$. Moreover, the oscillations were
observed at relatively high temperature, $T$, such that usual Shubnikov-de Haas
oscillations in the absence of the microwave irradiation were not
seen, $T\gtrsim \hbar\w_c$. Working with cleaner samples,
almost simultaneously,
two experimental groups~\cite{Mani,Zudov2} reported observations of
a novel zero-resistance state in two-dimensional electron systems,
appearing when the oscillations of the resistivity hit zero.
It is worth emphasizing that the zero-resistance state was not
connected to any significant features in the Hall resistivity in
contrast to that for the Quantum Hall Effect\cite{QHEreview}.
Further experimental activity consisted in analysis of
the low-field part of the oscillations in order to understand
the effect of the spin-orbit interaction\cite{Mani2} and observation
of the zero-conductance state in the Corbino disk geometry\cite{Zudov4}.
Results~\cite{Mani,Zudov2} were later confirmed
by an independent experiment.\cite{Dorozhkin}

Two recent theoretical papers\cite{Durst,Andreev}
are likely to explain the main qualitative features
of the data\cite{Zudov1,Mani,Zudov2,Mani2,Dorozhkin,Zudov4}.
Durst {\em et al.}\cite{Durst} presented a
physical picture and a calculation of the effect of microwave
radiation on the impurity scattering processes of a two dimensional
electron gas [see also Ref.~\onlinecite{Ryzhii1}].
In addition to obtaining big oscillations of the
magnetoresistance with the right period of Ref.~\onlinecite{Zudov1}.
the
crucial result of Refs.~\onlinecite{Ryzhii,Ryzhii1,Durst}
is the existence of the
regimes of magnetic field and applied microwave power for which the
longitudinal linear response conductivity is negative,
\begin{equation}
\sigma _{xx}<0.  \mylabel{eq0}
\end{equation}

It was shown by Andreev {\em et al.}\cite{Andreev}
that Eq.~(\ref{eq0}) by itself suffices to
explain the \emph{zero-$dc$-resistance} state observed in
Refs.~\onlinecite{Mani,Zudov2},
independent of the details of the microscopic
mechanism which gives rise to Eq.~(\ref{eq0}). The essence of
 Andreev {\em et al.}\cite{Andreev}
result is that a negative linear response conductance implies that the
zero current state is intrinsically unstable: the system spontaneously
develops a non-vanishing local current density, which almost
everywhere has a specific magnitude $j_0$ determined by the condition
that the component of electric field parallel to the local current
vanishes, see also Sec.~\ref{domain} of the present paper.
The existence of this instability was shown
 to be the origin of the observed zero resistance state. It is worth
mentioning that the instability of the system\cite{Gunn} with absolute
negative conductivity is known since the work of
Zakharov.\cite{Zakharov} The important new feature
of the instability and the domain structure of Andreev {\em et al.}
\cite{Andreev},
that the instability occurs at large Hall angle; as the result
the domains for the current coincide with the domains of the
electric field directed perpendicular to the current.
We also would like to notice the certain similarity with the model
of photoinduced domains proposed by D'yakonov\cite{Dyakonov}
as an explanation of the experiments on ruby crystals under
the intense laser irradiation\cite{Ruby}.

Subsequent theoretical works outlined ideas\cite{Anderson,HeandShe,Volkov}
of Ref.~\onlinecite{Durst,Andreev};
postulated\cite{Mikhailov} the plasma drift instability;
considered ``a simple classical model for the negative
dc conductivity'' due to non-parabolicity of the spectrum
\cite{newRaikh} or due to the lattice effects  on
ac-driven 2D electrons.\cite{Rivera}
We will not discuss those works further in a present paper.

Unfortunately, the comprehensive quantitative description of the
data\cite{Zudov1,Mani,Zudov2} is not possible within
Refs.~\onlinecite{Durst,Ryzhii,Ryzhii1}. Moreover, phenomenology
of Refs.~\onlinecite{Andreev} implies certain form of the
non-linear $dc$ transport in the presence of the microwave
radiation which has not been microscopically justified yet. Our
article presents a program for such a description. However, we
will not take into account effects which depend on the
distribution function and are determined by inelastic processes,
see Ref.~\cite{Mirlin-ac} and Section~\ref{sec:qshdH}. These
effects will be considered elsewhere.\cite{inelastic}

The paper is organized as follows. Qualitative discussion based on
a consideration of semiclassical periodic orbits is presented in
Sec.~\ref{qd}. The quantum Boltzmann equation applicable for large
filling factors and small angle scattering on the impurity
potential is derived in Sec.~\ref{qbe}. This equation is later
used to obtain closed analytic formulas for linear-$dc$ transport,
Sec.~\ref{Linear}, non-linear $dc$ transport, Sec.~\ref{nlinear},
and the effect of microwave radiation on the $dc$ transport,
Sec.~\ref{mwdc}. Section \ref{domain} relates the results to the
model of domains of Ref.~\onlinecite{Andreev}. Our findings are
summarized in Conclusions.

\section{Qualitative discussion}
\label{qd}

The qualitative discussion of the effect of the microwave
radiation on
the $dc$-transport was presented in
Refs. \onlinecite{Ryzhii,Ryzhii1,Durst}
in terms of {quantum}  transitions between Landau levels.
We chose to utilize the fact that only electrons with large
Landau level indices are important and explain the effects in terms of
semiclassical periodic motion. This explanation becomes
especially convenient
when the Landau levels are significantly broadened, which means that
the number of the repetitions in the periodic orbit is small.
[Infinite number of the repetitions of the periodic orbit would
correspond to the vanishing width of the Landau levels.]
Moreover, the qualitative picture will enable us to
separate effects into two groups according to their sensitivity
to the electron distribution function, and understand the status
and validity of the approximation which will be made in the technical
part of the paper.

To analyze the effect of external  fields on
the collision processes, it is more convenient to switch into the
 moving coordinate frame
\be
\r \to \r - \bzeta(t),
\mylabel{Galilean}
\ee
in which the external electric field is absent.
The position of the moving frame $\bzeta(t)$ is found from
\be
\pt \bzeta(t)=
\left(\frac{\pt -\w_c \hat{\varepsilon}}{\pt^2+\w_c^2}\right)
\frac{e\E(t)}{m_e},
\mylabel{zeta}
\ee
where $\E(t)$ is the applied spatially homogeneous electric field,
$m_e$ is the electron band mass,
$\w_c=\frac{eB}{m_ec}$
is the cyclotron frequency, and $\hat \varepsilon$ is the
antisymmetric tensor:
$\varepsilon_{xy}= -
 \varepsilon_{yx}= 1, \varepsilon_{xx}=\varepsilon_{yy}=0$.

If there were no disorder potential, the distribution function $f(\ep)$
of the electrons in this moving frame would be the Fermi function
\begin{equation}
f(\ep)=f_F(\ep)=\frac{1}{1+e^{\ep/T}},
\mylabel{FermiE}
\end{equation}
no excitations would appear and therefore no {dissipative} current
 would be possible. On the classical level,
an electron experiences the cyclotron motion with the position
of the guiding center $\R$ intact. Collision with
impurities  moving with velocity $\pt\bzeta$
cause  the drift of the guiding center, so that
the current density
\be
\j^{(d)}=eN_e\left(\frac{d\R}{dt}\right)
_{coll}
\mylabel{q1}
\ee
arises. Here, $N_e$ is the electron density, and
$\left(d\R/dt\right)_{coll}$ symbolizes the probabilistic
change in the position of the guiding center to be discussed below.

\begin{figure}
\epsfxsize=0.25\textwidth
\epsfbox{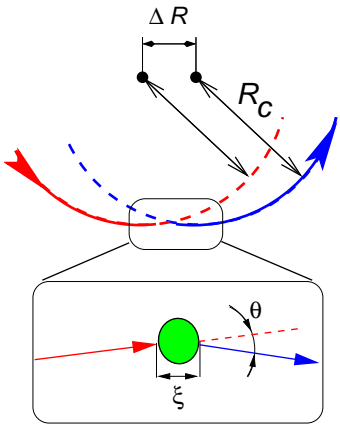}
\caption{Scattering process off a single
impurity. Inset shows the semiclassical trajectory in the vicinity
of the impurity. }
\label{fig:q1}
\end{figure}

Let us consider the scattering process of the electron off one
impurity. Because the size of the scattering region (correlation
length of the potential $\xi$) is much smaller than the cyclotron
radius, we can still characterize the scattering
process by the initial direction $\nn$, and the scattering angle
$\theta$ as shown on Fig.~\ref{fig:q1}.
 We consider only
small angle scattering
\be
\theta \ll 1.
\mylabel{q3}
\ee
In this case, each scattering event causes the shift in the
position of the guiding center
\be
\Delta \R = -\nn R_c \theta,
\mylabel{q4}
\ee
where $R_c=v_F/\w_c$ is the cyclotron radius and $v_F$ is the Fermi
velocity, see Fig.~\ref{fig:q1}.

During the collision process impurity moves with the velocity
$-\pt\bzeta(t)$. Because the size of the impurity is small, and we
assume that $|\pt\bzeta(t)| \ll v_F$, this motion
can be neglected in the calculation of the scattering
amplitude but it has to be taken into account in the conservation of
energy. Indeed, during the scattering event the moving
impurity transfers the energy
$\Delta\ep=-\pt\bzeta(t)\cdot\Delta{\p}$,
where change of the electron momentum is given by
\be
\Delta\p=-\hat\ep\nn p_F\theta
\mylabel{q5}
\ee
for $\theta \ll 1$, and $p_F=m_e v_F$ is the Fermi momentum.

Taking this energy change into account we write
for the  displacement of the center of orbit
\be
\begin{split}
&\frac{1}{R_c}\left(\frac{d\R}{dt}\right)
_{coll}\\
&=\Big\langle
\int d\theta
\frac{\Delta \R}{R_c}
\int d\ep
\left\{
f\left(\ep\right)-
f\left(\ep+\Delta\ep\right)
\right\}
 {\cal M}\Big\rangle_{\nn}
\\
&
=\Big\langle\nn
\int d\theta
\, \theta
\int d\ep
\frac{\partial f(\ep)}{\partial \ep}
\left(\pt\bzeta(t)\cdot\Delta{\p}\right) {\cal M}\Big\rangle_{\nn}
,
\end{split}
\mylabel{q6}
\ee
where the function
${\cal M}(\theta)$  is proportional to the scattering cross-section
and is determined by the impurity
potential. We will assume that ${\cal M}(\theta)$
vanishes rapidly at $\theta \sim \hbar/(p_F\xi) \ll 1$,
i.e. \req{q3} holds.
In \req{q6}, $\langle\dots\rangle_{\nn}$ stands for the averaging over the
direction of the momentum of the electron incoming on the impurity.

Substituting \req{q5} into \req{q6},
one finds
\be
\left(\frac{d\R}{dt}\right)
_{coll}=-\left(\frac{R_c}{\tau_{tr}}\right)
\left(\frac{\hat\ep\pt\bzeta}{v_F}\right),
\mylabel{q7}
\ee
where
\be
\frac{1}{\tau_{tr}}=\frac{p_Fv_F}{2}
\int d\theta \theta^2 {\cal M}(\theta)
\mylabel{qtautr}
\ee
is the transport scattering time at zero magnetic field. Together
with \req{q1} this gives the  Drude formula
$\sigma_{xx}=\frac{e^2N_e}{m\w_c^2\tau_{tr}}$ for the large Hall
angle $\w_c\tau_{tr}\gg 1$.

It is not the end of the story though. Considering one scattering
event as a complete real process, we imply that there are no
returns of an electron to the same impurity, or, to be more
precise, the possible returns are not correlated with original
scattering. However, in magnetic field an electron
moves along a circle of the cyclotron radius $R_c$ between
shattering processes. This circular motion results in correlated
returns of the electron to the same impurity.
[To the best of
our knowledge the first discussion
of the magnetoresistance in terms of returning semiclassical orbits was
performed in Ref.~\onlinecite{Baskin79}]

\begin{figure}
\epsfxsize=0.4\textwidth
\epsfbox{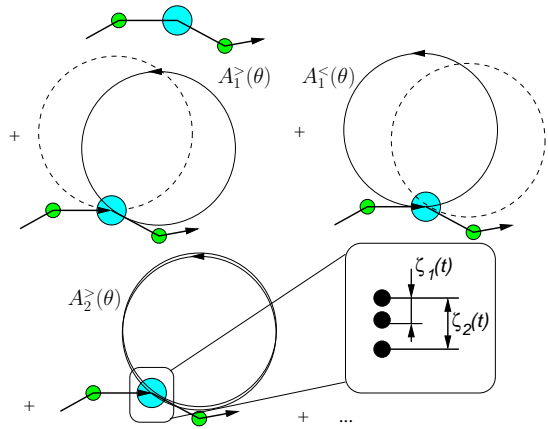}
\caption{
Different amplitude giving coherent contributions into
the impurity scattering cross-sections. Inset shows the shift
of the impurity between the scattering events
in moving coordinate frame.
}
\label{fig:q2}
\end{figure}

Such returns do not change the structure of \req{q6}, but they do
change the scattering cross-section ${\cal M}(\theta)$ in comparison
with its value in zero magnetic field. Indeed, one can see from
Fig.~\ref{fig:q2} that several semiclassical paths
characterized by different number of rotations and different instances
of the impurity scattering contribute to the same final
state; amplitudes for such processes $A_l^{\alpha}(\theta)$
sum up coherently.
\be
\begin{split}
{\cal M}(\theta) &\propto
\left|\sum_{l,\alpha}{A_l^{\alpha}(\theta)}\right|^2\\
&
=\left|{A_0(\theta)}\right|^2
+ 2\Re \sum_{
 l,l^\prime, \ \alpha\neq\alpha^\prime
}
{A_{l^\prime}^{\alpha^\prime}(\theta)}\left[A_{l}^\alpha(\theta)\right]^*
,
\end{split}
\mylabel{interf}
\ee
where index $l$ labels the number of rotation and $\alpha$
labels semiclassical paths; Fig.~\ref{fig:q2} shows
the paths for $l=1,2$, while Fig.~\ref{fig:q1} depicts the
path for $l=0$. Equation \rref{q6} takes into account
the contribution from the shortest trajectory (first term
in the second line) of \req{interf} and misses the
interference contributions.

To assess the role of the interference contributions, let us
employ the Born approximation of the impurity scattering.
Then, each semiclassical path may involve only one scattering
off an impurity, and all the paths are classified by
(i) the scattering angle $\theta$ and (ii) whether the impurity
affects the electron in the beginning or in
the end of the path; we will call corresponding amplitudes
$A^>_l(\theta)$ and $A^<_l(\theta)$, see
Fig.~\ref{fig:q2}.
Factorizing the impurity scattering potential into
the scattering cross-section at zero magnetic field
${\cal  M}_0(\theta)$,
we obtain
\be
{\cal M(\theta)}={\cal M}_0(\theta)
\left\{1+ 2\Re \sum_{
 l, l^\prime}
{A_{l^\prime}^>(\theta)}\left[A_{l}^<(\theta)\right]^*\right\}.
\mylabel{q8}
\ee

Equation \rref{q8} neglects the motion of the impurity
during the whole collision process. Apparently, it is
consistent with the derivation of \req{q6} where the
effect of the impurity motion on the matrix
elements was also neglected. Indeed, for the process
in Fig.~\ref{fig:q1}, $l = 0$, the characteristic
scattering time can be estimated as $\Delta t=\xi/v_F$.
The displacement of the impurity is, then,
$|\Delta\r|=|\pt \bzeta|\cdot(\xi/v_F) \ll \xi$, for $|\pt \bzeta| \ll v_F$,
and could be neglected. However,
for the scattering shown in Fig.~\ref{fig:q2} the interfering processes
are separated in time by the interval $\Delta t=\frac{2\pi l}{\w_c}$.
Displacement during such interval is
$\Delta|\r|\simeq |\pt \bzeta|\cdot\left(\frac{2\pi l}{\w_c}\right)
=|\pt \bzeta|\left(\frac{\xi}{v_F}\right)\left(\frac{2\pi l R_c}{\xi}
\right)$. Because $R_c\gg \xi$, the displacement may easily
become comparable with the impurity size $\xi$ even for $|\pt \bzeta|
\ll v_F$. Therefore the displacement must be taken into account in the
interference terms in \req{q8}. Due to the impurity motion, the
scattering off impurity occurs at different points,
see inset in Fig.~\ref{fig:q2}. In this respect,
the scattering off a moving
impurity is analogues to the interference in a grate
interferometer with
the distance between slits
equal to $\bzeta_l(t)=\bzeta\left(t-\frac{2\pi l}{\w_c}\right)
-\bzeta\left(t\right)$. By analogy with
the grate
interferometer, we find that \req{q8} has to be
modified as
\be
{\cal M}={\cal M}_0(\theta)
\left\{1+ 2\Re \sum_{
 l, l^\prime}
{A_{l^\prime}^>}\left[A_{l}^<\right]^*
\exp\left[\frac{i\Delta\p\cdot\bzeta_l\left(t\right)}{\hbar}
\right]
\right\},
\mylabel{q9}
\ee
where $\Delta\p$ is given by \req{q5}.
Finally, the quantum mechanical amplitudes $A_l^{>,<}(\theta)$
can be decomposed into the smooth prefactor  $a$ and the rapidly
oscillating exponent involving the reduced action along a semiclassical
cyclotron trajectory
\be
A_{l}^{>,<}=a_{l}^{>,<}
\exp\left[il\left(\frac{\pi p_F^2\lH^2}{\hbar^2}
 + \frac{2\pi\ep}{\hbar\w_c}\right)\right],
\mylabel{q10}
\ee
with $\lH=\left(\frac{\hbar c}{eB}\right)^{1/2}$ being the magnetic length.

Having investigated the effect of the returning paths on
the scattering process, we are ready to write an expression
for the dissipative current. Substituting \reqs{q9} and \rref{q10}
into \req{q6}, we find
\begin{widetext}
\be
\begin{split}
\left(\frac{d\R}{dt}\right)
_{coll}
&=R_c\Bigg\langle\nn
\int d\theta
\, \theta {\cal M}_0(\theta)
\int d\ep
\frac{\partial f(\ep)}{\partial \ep}
\left(\pt\bzeta(t)\cdot\Delta{\p}\right)
\\
&\times
\left\{1+ 2\Re \sum_{
 l, l^\prime}
{a_{l^\prime}^>}\left[a_{l}^<\right]^*
\exp\left(\frac{i\Delta\p\cdot\bzeta_l\left(t\right)}{\hbar}
+\frac{i\pi(l^\prime-l) p_F^2\lH^2}{\hbar^2} + \frac{i
\pi(l^\prime-l)\ep}{\hbar\w_c} \right) \right\} \Bigg\rangle_{\nn}
,
\end{split}
\mylabel{q11}
\ee
\end{widetext}
Equation \rref{q11} is the main qualitative result of this section
\footnote{Time argument in the work committed by impurity
$\left(\pt\bzeta(t)\cdot\Delta{\p}\right)$
also shifts by multiples of $2\pi/\w_c$, however, it
will not be important for the qualitative arguments,
see Sec.~\ref{qbe} for the rigorous theory.
}.
It shows, that due to the presence of the returning orbits
on one hand, and the largeness of the period on the other hand,
the scattering process is extremely sensitive to external fields
applied to the system. As the result, a rich variety
of the effects arises.
The effects can be separated into two groups: (i) sensitive
to the distribution function; and (ii) non-sensitive to
the electron distribution. We will discuss those groups
separately in  following two subsections.

\subsection{Effects dependent on the form of distribution function}
\label{sec:qshdH}

Retaining only the contributions from the amplitudes with different
winding numbers  in \req{q11}, and considering
only linear terms in $\bzeta$,
we obtain familiar Shubnikov -- de Haas oscillations
$\rho_{osc}$ of the resistivity $\rho_{xx}(B)$:
\be
\frac{\rho_{osc}(B,T)}{\rho_{xx}(B=0)}
=\sum_{m=1}
\eta_m
\Upsilon_m \cos\left(\frac{\pi m p_F^2\lH^2}{\hbar^2}\right)
\mylabel{q12}
\ee
where
\be
\Upsilon_m=-\int d\ep
\frac{\partial f(\ep)}{\partial \ep}
\exp\left[\frac{i 2\pi m\ep}{\hbar\w_c}\right]
\mylabel{q13}
\ee
The form-factors,
$\eta_m\simeq\sum_{l}a_l^<[a_{l+m}^>]^*
$ are determined by the impurities
at which scattering may occur during the circular motion of the
electron,
see Sec.~\ref{sec:qformfactors} for a more detailed discussion.
The accurate expression for the factors $\eta_m$ are written
in Sec.~\ref{sec:Shubnikov}.

Neglecting the effect of the applied electric field on the electron
distribution function $f(\ep)$, {\it i.e.} $f(\ep)=f_F(\ep)$ see
Eq.~(\ref{FermiE}), we obtain
\be
\Upsilon_m
=\frac{2\pi^2 m T}{\hbar\w_c \sinh \left(\frac{2\pi^2 m T}{\hbar\w_c}
\right)}.
\label{q13a}
\ee
Therefore, the Shubnikov -- de Haas oscillations are exponentially
suppressed at high temperature $T\gtrsim \w_c/\pi^2$.

The consideration of effects non-linear in the applied
electric field is more complicated task.
Indeed, electric field may significantly change the distribution
function $f(\ep)$. In particular, the shape of the distribution
function in a strong electric field may have nothing to
do with the Fermi distribution.
Because these effects are extremely sensitive to the form of
distribution function, their description
requires specifying the microscopic mechanism of the
energy relaxation.

We notice however that the smooth part of the
electron distribution function with the characteristic width
$T_{eff}\gtrsim T$ much larger than $\w_c$ results in
exponentially small contribution to the resistivity.
Therefore we focus our discussion to the high temperature
limit $T \gg \hbar\w_c$ and neglect the Shubnikov-de Haas
contribution to the non-linear effects in further consideration.
Unfortunately, the latter restriction does not allow us to avoid
consideration of non-linear effects of the electric field
on the electron distribution function.
Indeed, non-equilibrium component of the electron distribution
function produced by the impurity scattering oscillates with
period $\hbar\w_c$.\cite{Mirlin-ac} Substitution of such
oscillating function into Eq.~(\ref{q13}) results in a
contribution which is not exponentially small.
An estimate\cite{Mirlin-ac} shows that
this contribution may dominate all other effects, discussed below.
A more detailed analysis of the non-linear effects on the electron
distribution function and electron transport will be presented
elsewhere.\cite{inelastic}

\subsection{
Effects independent from the form of distribution function}
\label{sec:qsmooth}

The contribution which survive the thermal  or energy
 averaging
are only those with the same winding number $l=l'$.
Retaining only such terms in \req{q11}, we can see that the
energy dependence of the scattering cross-section vanishes,
so that the energy integral can be  evaluated.
The result does not depend on the distribution function
anymore:
\begin{widetext}
\be
\begin{split}
\left(\frac{d\R}{dt}\right)
_{coll}
=-R_c\Bigg\langle\nn
\int d\theta
\, \theta {\cal M}_0(\theta)
\left(\pt\bzeta(t)\cdot\Delta{\p}\right)
\left\{1+ 2\Re \sum_{l=1}^{\infty}
{a_{l}^>}\left[a_{l}^<\right]^*
\exp\left(\frac{i\Delta\p\cdot\bzeta_l\left(t\right)}{\hbar}
\right)
\right\}
\Bigg\rangle_{\nn}
.
\end{split}
\mylabel{q14}
\ee
\end{widetext}
Therefore, all of the non-linear phenomena come from the
effect of the external fields on the scattering cross-section
itself.

We consider the system under the effect of the $dc$ electric field
only. In this case
\be
\bzeta_l(t)=-\frac{2\pi l}{\w_c}\pt \bzeta;
\mylabel{q15}
\ee
In the linear response  $|\Delta \p\bzeta_l| \ll \hbar$,
we obtain that
the scattering rate is enhanced due to multiple returns:
\be
\frac{\rho_{xx}(B)}{\rho_{xx}(B=0)}
=1+2\Re\sum_{l=1}^\infty a^>_l\left[a^<_l\right]^*.
\mylabel{q16}
\ee
Postponing the estimate of the form-factors $a^>_l[a^<_l]^*>0$ until
subsection \ref{sec:qformfactors},
see also Sec.~\ref{Linear},
we notice that \req{q16} describes positive magnetoresistance.
Indeed, it is intuitively clear that stronger the magnetic field,
the larger the probability for electron to return to the same
impurity. Thus, the contribution of the sum in \req{q16} increases.

One can see that the contribution of the periodic orbit
produces non-linear current voltage characteristics.
Indeed, with the increase of the electric field $|\Delta \p\bzeta_l|/
\hbar$ becomes of the order of unity. Contribution of the returns
with corresponding winding number $l$ would be suppressed;
electron after $l$ turns simply
misses impurity. If  field is such $|\Delta \p\bzeta_{l=1}|/\hbar
\simeq 1$, then the contribution of all returns will be
suppressed. As the result, at $dc$ fields larger
than some value $E_0$ the current voltage characteristics becomes
linear but with the slope determined by the transport time
in the absence of magnetic field, see \req{qtautr}.
We estimate the value of $E_0$ by noticing that according to
\req{zeta}, $|\pt\bzeta|=eE_0/m\w_c$, and $|\bzeta_1|=2\pi
E_0/m_e\w^2_c$. Then, using $|\Delta \p| \simeq
\hbar/\xi$, where $\xi$ is the correlation radius of the impurity
potential, we obtain the characteristic field
\be
eE_0\simeq m\w_c^2\xi
\mylabel{q17}
\ee
Accurate theory of nonlinear effects in the $dc$ field is contained in
Sec.~\ref{nlinear}.

Assume now that the $ac$ microwave field with the frequency
$\w$ is applied together
with the $dc$ field. The velocity of electrons due
to those fields $\pt\bzeta(t)$
and the displacement during $l$ periods $\bzeta_l(t)=
\bzeta(t-2\pi l/\w_c)-\bzeta(t)$ can be found as
\be
\begin{split}
&\pt\bzeta(t) =\v_{dc} + \w \bzeta_{ac}\cos\w t,\\
&\bzeta_l(t)=-\frac{2\pi l\v_{dc}}{\w_c}-{2 \bzeta_{ac}}
\sin\left(\frac{\pi l \w}{\w_c}\right)\cos\left(\w t
-\frac{\pi l\w}{\w_c}
\right).
\end{split}
\mylabel{q18}
\ee
In the linear response regime those two velocities give two
independent contributions to the current, i.e. $dc$ response is not
affected by $ac$ radiation at all. Presence of the non-linear term in
the collision probability \req{q14} results in the photovoltaic
effect.\cite{Belinicher}
Indeed, expanding the exponent in \req{q14} up to the second
order and averaging the result over time, we estimate
\be
\begin{split}
&\Big\langle \pt\bzeta(t)
\exp\left(\frac{i\Delta\p\cdot\bzeta_l\left(t\right)}{\hbar}
\right) \Big\rangle_t
\\
&\approx
\Big\langle \pt\bzeta(t)
\left[1-\frac{1}{2\hbar^2}
\left(\Delta\p\cdot\bzeta_l\left(t\right)
\right)^2 \right]\Big\rangle_t
\\
&
\approx
\v_d\left[1-
\frac{\Delta\p^2\bzeta_{ac}^2\sin^2\frac{\pi l\w}{\w_c}}{\hbar^2}
- \frac{2\pi l\w}{\w_c}
\frac{\Delta\p^2\bzeta_{ac}^2\sin\frac{2\pi l\w}{\w_c}}{\hbar^2}
\right].
\end{split}
\mylabel{q19}
\ee
It is important to emphasize, that at the frequency
of $ac$ field commensurate with the cyclotron frequency,
$\w=j\w_c$, the $ac$ field does not affect the $dc$ resistivity.
This result is easy to understand by noticing that under this
condition the impurity returns to its initial position during
one cyclotron period.

The last term in \req{q19} represents the photovoltaic effect and
deserves some attention.  Substituting
\req{q19} in \req{q14} and neglecting the second term in \req{q19}, we obtain
\be
\frac{\rho_{xx}(B)}{\rho_{xx}(0)}
\approx 1+2\Re\sum_{l=1}^\infty
\left(1-\frac{2\kappa_l \pi l \w}{\w_c}\sin\frac{2\pi l\w}{\w_c}\right)
a^>_l\left[a^<_l\right]^*,
\mylabel{qphvolt}
\ee
where $\kappa_{l} \propto |\bzeta_{ac}|^2$.
The physical meaning of the photovoltaic term is the
rectification of the $ac$ current due to the non-linear
term in the collision integral. In the absence of the $dc$
field there is no preferred direction, so that the rectified
current vanishes, whereas application of the $dc$ voltage
defines this direction.
One can see  that, due to the large factor $2\pi\w/\w_c \gg 1$,
this term can exceed unity
even if $|\Delta p\bzeta_a| \lesssim \hbar$. Therefore, the
sign of the contribution from the returning orbits
may be changed in comparison with the $dc$ result,
compare  \reqs{q16} and \rref{qphvolt}.
The zero-voltage $dc$ resistivity may become negative.
Accurate results for the $dc$ response in the presence of the microwave
are collected in Sec.~\ref{mwdc}.

Last comment concerns microscopic justification of
the main assumption of Ref.~\onlinecite{Andreev}, that
even if the zero $dc$-current resistivity under microwave
radiation is negative, it becomes positive
at large enough applied $dc$ current. Indeed, according
to the arguments before \req{q17}, the contribution of the
cyclotron orbits vanishes if the electric field is large enough.
On the other
hand, these are the only contributions affected by the microwave radiation.
Thus, at   applied $dc$ electric field exceeding
$E_0$ the current voltage characteristics becomes
linear, with the slope determined by the transport time
in the absence of magnetic field \req{qtautr},
and not sensitive to the effect of
the microwave.

\subsection{Form-factors, self-consistent Born approximation, and
classical memory effects}
\label{sec:qformfactors}

Equation \rref{q14} derived on quite general grounds would describe all
of the physics quantitatively if the finding of the form-factors
$a_l^{<,>}$ were a trivial task. Unfortunately, it is not so,
and the quantitative analysis of the transport requires a
machinery developed in the subsequent sections. The purpose
of the present subsection is to explain the structure of
the form factors qualitatively, and to clarify the physical
meaning and the status of the approximations to be employed.

\begin{figure}
\epsfxsize=0.4\textwidth
\epsfbox{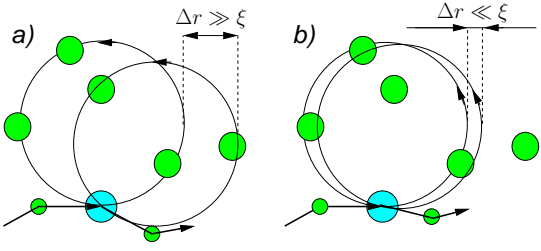}
\caption{
a) Pair of typical
semiclassical trajectories
and; b) of those
contributing to the classical memory
effect.
}
\label{fig:q3}
\end{figure}

Let us start from the contribution of the returning paths
with one turn, $a^{<,>}_1$. Those factors are
disorder specific quantities and we have to average them over the
disorder realizations. The estimate for such averaging proceeds
as following.
Let us assume that the two trajectories travel through
the different impurities as shown in Fig.~\ref{fig:q3}a.
Then, every impurity scattering will randomize the sign
of  $a^{<}_1a^{>}_1$. Therefore, the only remaining contribution
originates from trajectories which were not affected by other
impurities during the cyclotron motion at all.
The amplitudes can be estimated from
$|a^{<,>}_1|^2=P_0\left(\frac{2\pi}{\w_c}\right)$,
where $P_0(t)=\exp(-t/\tau_q)$ is the probability
for electron not to be scattered at an angle during time $t$,
and
\be
\frac{1}{\tau_q}=p_Fv_F\int d\theta M(\theta) \simeq \frac{p_F^2\xi^2}
{\hbar^2\tau_{tr}} \gg \frac{1}{\tau_{tr}}.
\mylabel{qtauq}
\ee
is the quantum lifetime.
This gives the estimate (which is actually an exact answer)
\be
a^{<}_1=a^{>}_1=e^{-\frac{\pi}{\w_c\tau_q}}.
\mylabel{q20}
\ee

One may naively  try to estimate the contribution from the paths with
two turns from
\[
[a^{<}_2]^2\stackrel{?}{=}P_0\left(2\cdot\frac{2\pi}{\w_c}\right);
\]
which leads to $a^{<}_2=a^{>}_2\stackrel{?}{=}e^{-\frac{2\pi}{\w_c\tau_q}}$
However, this estimate is not correct.
The reason is that the path which was scattered off an
impurity on the first turn can be rescattered off the same
impurity once again and give the contribution which is not
random, see Fig. \ref{fig:q3}a. In the absence of the
external electric field or the microwave illumination
it gives
\be
a^{<}_2=a^{>}_2=
\left(1- \frac{2\pi}{\w_c\tau_q}\right)
e^{-\frac{2\pi}{\w_c\tau_q}},
\mylabel{q21}
\ee
which differs from the naively expected value.
Moreover, it is clear that
$a_2$ will be affected by the motion of the impurity,
i.e. the formfactors by itself are  functions of the $dc$ and
microwave field [this effect was discussed from a different
point of view in Ref.~\onlinecite{Ryzhii2}].

\begin{figure}
\epsfxsize=0.4\textwidth
\epsfbox{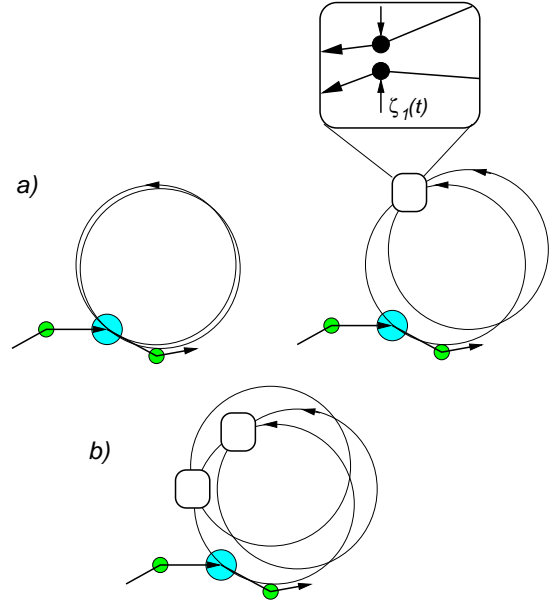}
\caption{
Trajectories with (a) two and (b) three turns contributing to the
form-factors $a_{2,3}^>$; Inset: the semiclassical path
of the electron in the vicinity of the impurity.
}
\label{fig:q4}
\end{figure}

One can see, e.g. from Fig.~\ref{fig:q4}b, that the larger the number
of turns, the larger the number of the returning orbits must be taken into
account. To perform analytical calculations, we utilize
the self-consistent Born approximation (SCBA)
(first used for the disordered electrons in the magnetic field by
Ando and Uemura\cite{Ando}). This approximation enables us to
describe the combinatorial
factors and the effect of the fields on the intermediate scattering
processes correctly.

The drawback of the SCBA is that
the disorder potentials
acting on electron on different semiclassical paths are assumed to
be uncorrelated with each other.
The justification of such an approximation  requires
that the typical distance between trajectories, see
Fig. \ref{fig:q3}a, to be larger than the correlation radius of the
potential $\xi$. On the other hand this distance can be estimated
as $R_c\theta\simeq \hbar R_c/(p_F\xi)$. We obtain
the inequality $\xi \ll \hbar R_c/(p_F\xi)$
or
\be
\xi^2 \ll \frac{\hbar R_c}{p_F}=\lH^2,
\mylabel{eqRaikh}
\ee
where $\lH$ is the magnetic length. Condition of the
validity of the self-consistent Born approximation was first
established by Raikh and  Shahbazyan\cite{Raikh} by an explicit
calculation of the first
correction to the self-consistent Born approximation.

The case of weak magnetic field such that only single turn
trajectories remain $1/\tau_{tr} \ll \w_c \ll 1/\tau_q$,
requires additional consideration, even if the criteria
\rref{eqRaikh} is satisfied.
The trajectories close to each other, shown in Fig.~\ref{fig:q3}b,
give the main contribution to the interference of return
trajectories, even though the fraction of these non-typical
trajectories is small.
Indeed, these trajectories travel through the same disorder.
The scattering off
the same disorder does not randomize the sign of the
product $\Re(a^{<}_1a^{>*}_1)$, and thus exponential estimate no longer
holds. On the contrary,  for $\theta \lesssim \frac{\hbar}{p_F\xi}$,
one can roughly estimate from Fig.~\ref{fig:q3}b
\[
a^{<}_1(\theta)a^{>}_1(\theta) \simeq \exp\left[
-\frac{2\pi}{\w_c\tau_q}\left(\frac{\Delta \r}{\xi}\right)^2\right],
\]
where $\Delta\r$ is the typical distance between
trajectories. Estimating $|\Delta \r| \simeq R_c\theta_1$, one
finds for the angular interval of non-typical trajectories
\[
\begin{split}
\theta_1&\simeq \frac{\xi}{R_c}\left(\w_c\tau_q\right)^{1/2}
= \left(\frac{\hbar}{p_F\xi}\right) \left(\frac{\xi}{R_c}\right)
\left(\frac{\w_c\tau_q p_F^2\xi^2}{\hbar^2}\right)^{1/2}
\\
&
=  \left(\frac{\hbar}{p_F\xi}\right) \left(\frac{\xi}{R_c}\right)
\left(\w_c\tau_{tr} \right)^{1/2}.
\end{split}
\]
The contribution of the scattering process
to the transport scattering time \rref{qtautr}, is proportional
to $\theta_1^3$ and the typical scattering angle is $\hbar/(p_F\xi)$.
Therefore the relative contribution to the non-typical
trajectories to the change in all the dissipative processes,
say resistivity, is
\be
\frac{\Delta \rho_{xx}}{\rho_{xx}} \simeq
\left(\frac{p_F\xi\theta_1}{\hbar}\right)^3 = \left(\frac{\xi}{R_c}\right)^3
\left(\w_c\tau_{tr} \right)^{3/2}.
\mylabel{q22}
\ee
This power law dependence should replace
$\exp(-\frac{2\pi}{\w_c\tau_q})$
dependence in all the formulas obtained in
the  self-consistent Born approximation for $\hbar\w_c\gg T$ (not
containing Shubnikov - de Haas oscillatory term).

Finally, we notice that the product of two  amplitudes for
the electron propagation in the same disorder can be described as the
classical probability of the circular path, and
all the discussion can be recast into the notion of the classical
memory effect (CME). It is not accidental
that the estimate \rref{q22} coincides with calculation of the
CME magnetoresistance \cite{mirlin1} up to numerical
prefactor.
Accurate calculation of the non-linear effect within CME model
will not be done in the present paper.
It is important to emphasize, however, that the fact
that the self-consistent Born approximation has to
be corrected by classical memory effect, affects only overall prefactors
in the non-linear effects and do not change the basic structure
of \req{q14}.\cite{memoryeffect}

\section{Quantum Boltzmann equation (QBE)}
\label{qbe}

In this section, we will derive the quantum Boltzmann
equation. The propose of this derivation is to
separate the contributions remaining at $\nu \gg 1$ from the very
beginning.
We use the standard
Keldysh formalism for the non-equilibrium system\cite{Keldysh}.

\subsection{Derivation of the semiclassical transport equation}

The derivation is very similar to that for the
Eilenberger equation \cite{Eilenberger,68}.
The matrix Green functions and the corresponding self-energies
have the form
\be
\hat{G}=\begin{pmatrix}
\hat G^R & \hat G^K \\ 0 & \hat G^A
\end{pmatrix}_K;
\quad
\hat{\Sigma}=\begin{pmatrix}
\hat\Sigma^R & \hat\Sigma^K \\ 0 & \hat\Sigma^A
\end{pmatrix}_K,
\mylabel{Gr1}
\ee
where the component of the matrices are linear
operators in time space and the one electron Hilbert space.

The equation for the Green function is
\be
\left(i\pt -\hat{H}\right)\hat{G}=\openone+\hat{\Sigma}\hat{G};
\quad  \hat{G}\left(i\pt -\hat{H}\right)=\openone+\hat{G}\hat{\Sigma}.
\mylabel{Gr2}
\ee
where $\hat{H}$ is the one-electron Hamiltonian of the clean system,
$\openone$ is the short hand notation $\openone=I_K\otimes I_e
\delta\left(t_1-t_2\right)$ and $I_{K,e}$ are the unit matrices in the
Keldysh space and the one electron Hilbert space respectively.
[We  put $\hbar=1$ in all of the intermediate formulas.]
For retarded Green function we use
\be
\left(i\pt -\hat{H}\right)
\hat{G}^R= \delta(t-t_1)\hat I_e +\hat\Sigma^R\hat{G}^R;
\quad \hat G^R\left(t< t_1\right)=0.
\mylabel{Gr3}
\ee
and $\hat{G}^A=\left[\hat{G}^R\right]^\dagger$.
For the Keldysh Green function it is convenient to take the
non-diagonal component of the difference of two \reqs{Gr2}
\be
\left[\left(i\pt -\hat{H}\right); \hat{G}^K
\right]
= \hat\Sigma^R\hat{G}^K - \hat{G}^K
\hat\Sigma^A +\hat\Sigma^K{G}^A - \hat G^R\hat \Sigma^K,
\mylabel{Gr4}
\ee
where $[\cdot;\cdot]$ stands for the commutator.
The next standard step is to separate the time evolution of the occupation
numbers $\hat f$ and the wave-function of the system
\be
\hat{G}^K=\hat G^R-\hat G^A - 2
\left[\hat G^R\hat{f} - \hat{f} \hat G^A\right].
\mylabel{Gr5}
\ee
In general  $\hat{f}$ is an operator in both one-electron space and
the time space. In the thermal equilibrium, however, one has simply
\be
\begin{split}
&\hat{f}_{F}=
\int \frac{d\ep}{2\pi}e^{-i\ep(t_1-t_2)}
f_F(\ep) =\frac{iT}{2\sinh \pi T(t_1-t_2+i0)}, \\
&  f_F(\ep)=\frac{1}{e^{\ep/T}+1},
\end{split}
\mylabel{Gr5a}
\ee
where $T$ is the temperature in the energy units.
Substituting \req{Gr5} into \req{Gr4}, one obtains the kinetic
equation
\begin{subequations}
\label{keq1}
\be
\left[\left(\pt + i \hat{H}\right); \hat f
\right]= \hat\St\{\hat f\},
\mylabel{Gr10}
\ee
with the collision integral given by
\be
i\hat \St\{\hat f\} \equiv
\left[\hat \Sigma^R \hat f -\hat f\Sigma^A\right]
+\frac{1}{2}\left[\hat\Sigma^K - \hat\Sigma^R+\hat\Sigma^A\right].
\mylabel{Gr10a}
\ee
\end{subequations}

\begin{figure}
\epsfxsize=0.45\textwidth
\epsfbox{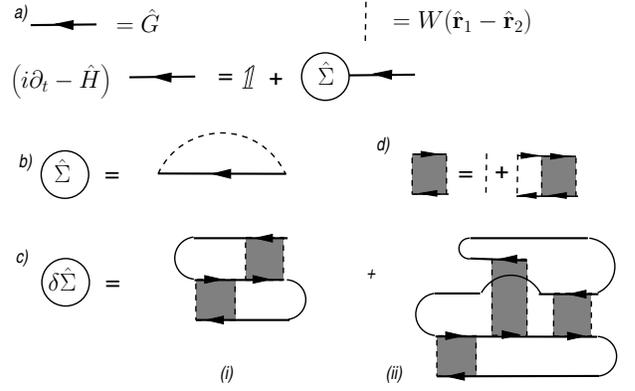}
\caption{(a)-(b) Self-consistent Born approximation (SCBA).
Diagrams (c)  are most important
contributions not included in the SCBA, corresponding
to the logarithmically divergent second loop weak localization
correction\cite{Wegner79}, see {\em e.g.} Ref.~\onlinecite{Benedict}.
Diagram (c (i)) with one of the diffusons (d) reduced to
one impurity line describes the first contribution
to the classical memory magnetoresistance\cite{mirlin1}.
}
\label{fig:qbe1}
\end{figure}

Next step is to write down a self-energy for the electron subjected
to the random potential $U(\r)$ characterized by the correlation
function
\[
\langle U(\r_1) U(\r_2) \rangle= \int \frac{d^2q}{(2\pi)^2}
W(q)e^{i\q(\r_1-\r_2)}.
\]
For the sake of concreteness we will adopt the model with
\be
W(q)=W(0)e^{-q\xi}
\mylabel{disorder}
\ee
where $\xi$ is the disorder correlation length.
Equation \rref{disorder} is an adequate description for
the potential created by remote donors situated on the distance
$\xi/2$ from the plane of two-dimensional electron gas.
Self-consistent Born approximation, involving the summation
of all the diagrams with non-intersecting impurity lines,
see Fig.~\ref{fig:qbe1} is
\be
\hat\Sigma=
 \int \frac{d^2q}{(2\pi)^2}
W(q)\left[e^{i\q\hat\r}
\hat G e^{-i\q\hat\r}\right].
\mylabel{SCBA}
\ee

Self-consistent Born approximation is justified if two conditions
\begin{subequations}
\label{SCBAvalid}
\bea
\xi \ll \lH, \label{SCBAvalida} \\
\sigma_{xx} \gg \frac{e^2}{2\pi\hbar}
\label{SCBAvalidb}
\eea
hold.
Hereinafter $\lambda_H=\left(\frac{c\hbar}{eB}\right)^{1/2}$
is the magnetic length, and $B$ is the applied magnetic field.
The physical meaning of the condition \rref{SCBAvalida} was
discussed in Sec.~\ref{qd}, see \req{eqRaikh}.
Condition \rref{SCBAvalidb}
allows us to neglect the localization effects shown on
Fig.~\ref{fig:qbe1}c.
Note, that the external microwave radiation further suppresses
the localization correction\cite{Altshuler}.
\end{subequations}

We also assume small angle scattering
\be
p_F\xi \gg 1,
\mylabel{smallangle}
\ee
where $p_F$ is the Fermi momentum. This condition
is not really essential for the physical processes
but it allows for some technical simplifications.
\footnote{Reliable estimate for the short range scatterer
can be obtained by substitution $\xi \to \hbar/p_F,
\ \tau_q=\tau_{tr}$ in the final formulas.} Moreover,
based on Shubnikov de-Haas data,
we believe that this regime is the most relevant for the
experimental situation.\cite{Zudov1,Mani,Zudov2}

We will consider the system in the classically
strong magnetic field, so that the Hall angle $\w_c\tau_{tr}$ is large
($\tau_{tr}$ is the transport time, see \req{tautr} below).
The effects we will be studying are proportional to the
inverse Hall angle and vanish in clean systems.
Therefore, it is convenient to solve time dependent problem for the
clean system first, and then to consider the effect of the disorder on the
top of this solution. This program is easily accomplished by using the
transformation \reqs{Galilean} and \rref{zeta}.
Transformation \rref{Galilean} removes electric field from the
Hamiltonian. Instead, the
disorder potential becomes time dependent.
We rewrite \req{SCBA} as
\be
\begin{split}
&\hat\Sigma=
 \int \frac{d^2q}{(2\pi)^2}
W(q)e^{-i\q\bzeta_{12}}
\left[e^{i\q\hat\r}
\hat G e^{-i\q\hat\r}\right];
\\
&
\bzeta_{12}=\bzeta(t_1) - \bzeta(t_2).
\end{split}
\mylabel{SCBAt}
\ee

Separating the guiding
center coordinate and the cyclotron motion, we write
\be
\hat{H}=\frac{\hat \p^2}{2m_e} - \mu;
\quad \hat \r=\hat{\R}+\lH^2\hat \ep\hat{\p},
\mylabel{H}
\ee
where $\mu$ is the chemical potential and the operators obey
the following commutation relations
\be
\mylabel{Commutators}
 \left[\hat{R}_\alpha; \hat{R}_\beta\right]=i\lH^2\ep_{\alpha\beta};
\ \left[\hat{p}_\alpha;
  \hat{p}_\beta\right]=-\frac{i}{\lH^2}\ep_{\alpha\beta};
\ \left[\hat{\R};
  \hat{\p}\right]= 0.
\ee
The Green functions $\hat G(\hat \p;\hat \R)$ and
the self-energies  $\hat \Sigma(\hat \p;\hat \R)$
can be obviously written
as functions of the operators \rref{Commutators}.
Using commutation relations \rref{Commutators},
we obtain from \req{SCBAt}
\be
\begin{split}
 &\hat\Sigma
 \left(\hat\p;{\hat\R}\right)
 =
 \int \frac{d^2q}{(2\pi)^2} W_{12}
\left[e^{i{\bf q}\ep\hat{\bf p}\lH^2}
\hat G\left(\hat\p;{\hat\R}(\q)\right)
e^{-i{\bf
q}\ep\hat{\bf p}\lH^2}
\right]
\\
 & W_{12}\equiv W(q)e^{-i\q\bzeta_{12}};
\quad
{\hat\R}(\q)=
{\hat\R}+\lH^2\hat\ep\q.
\end{split}
\mylabel{LSCBAt}
\ee
Here we suppressed time indices.

Next step is to separate the  motion in the phase space
into components parallel and perpendicular to
the  Fermi surface. For this purpose, we parameterize
the cyclotron motion operators as
\be
\begin{split}
&\hat p_x=
\frac{1}{\lH}
\left(
\frac{{N+\hat n}}{{2}}
\right)^{1/2}
e^{i\hat\varphi} + h.c.;
\\
&\hat p_y=
\frac{-i}{\lH}
\left(
\frac{{N+\hat n}}{{2}}
\right)^{1/2}
e^{i\hat\varphi} + h.c,
\end{split}
\mylabel{parameter}
\ee
where the integer
\[
N={\rm Int}\frac{\mu}{\w_c}
\]
 is introduced for convenience.
To preserve the commutation relations \rref{Commutators}
for operators \rref{parameter}, the commutation relation
\be
\left[\hat n; \hat\varphi \right]=-i
\mylabel{Com2}
\ee
is imposed.
It follows from \reqs{Com2} and \rref{parameter} that the
integer eigenvalues $n\geq -N$
of the operator $\hat n$ have the meaning of the Landau level indices.
The Hamiltonian \rref{H} acquires the form
\be
\hat{H}=\w_c\left[\hat n + \delta(\mu)\right];
\
\delta(\mu)=\frac{1}{2} + N - \frac{\mu}{\w_c}.
\mylabel{H2}
\ee

All the previous manipulations were valid for any
magnetic field. Now we are going to make use of the large filling factor
\begin{subequations}
\label{Nbig}
\be
N \gg 1.
\mylabel{Nbiga}
\ee
We will assume that the characteristic value of $\hat n$ contributing
to the transport quantities is such that $\parallel\hat n \parallel
\ll N$, i.e. all the relevant dynamics occurs in the vicinity of the
Fermi level. This assumption is justified provided that two conditions
\be
T \ll N\hbar\w_c; \quad \w_c\tau_q \ll N
\mylabel{Nbigb}
\ee
are satisfied, with $\tau_q$ being the quantum elastic scattering time, see
below.
Those assumptions allow for the semiclassical consideration of
the self-energy \rref{SCBAt}, which is presented below.
\end{subequations}

Using \reqs{Nbig}, we expand  \req{parameter}
as
\be
\begin{split}
&\hat p_x= p_F\cos\hat\varphi
+ \frac{1}{2R_c}
\left[\hat n e^{i\hat\varphi}+h.c.
\right]
+\dots
;
\\
&\hat p_y=p_F\sin\hat\varphi
-  \frac{1}{2R_c}
\left[i\hat n e^{i\hat\varphi}+h.c.
\right]
+\dots,
\end{split}
\mylabel{parexpand}
\ee
where
\[
p_F=\frac{\sqrt{2N}}{\lH}; \quad R_c={\sqrt{2N}\lH}
\]
are the Fermi momentum and the cyclotron radius respectively.
Substituting \req{parexpand} into \req{LSCBAt} and
keeping in mind condition \rref{smallangle}, we find


\be
\begin{split}
 &\hat\Sigma
 \left(\hat n;\hat\varphi;{\hat\R}\right)
 =
 \int \frac{d^2qW_{12}}{(2\pi)^2}
\left[\hat U_\parallel
\hat U_\perp \hat G\left(\hat n; \hat\varphi; {\hat\R}(\q)\right)
\hat U_\perp^\dagger
\hat U_\parallel^\dagger
\right],\\
&\hat U_\perp = \exp\left[iqR_c\sin \hat\varphi_q
-\frac{iq^2R_c
\sin 2\hat\varphi_q
}{4p_F} + {\cal O}\left(\frac{q^3R_c}{p_F^2}\right)\right],
\\
&\hat U_\parallel  = \exp\left\{\frac{q
\left[\hat{n}e^{i\hat\varphi_q} - h.c.\right]
}{2p_F}
\right\}, \quad \hat\varphi_q\equiv\hat\varphi-\phi_q,
\end{split}
\mylabel{LSCBA1}
\ee
where $\q=q(\cos\phi_q; \sin\phi_q)$.
Matrix $\hat U_\perp$ commutes with $\hat\varphi$, {\em i.e.}
in semiclassical sense
it describes the scattering of the electron perpendicular to
the Fermi surface. On the other hand,  matrix $\hat U_\parallel$
changes $\hat\varphi$, i.e. it has the semiclassical meaning of
the evolution parallel to the Fermi surface.
Due to the condition \rref{Nbiga} those processes originate from
parametrically different values of $\q$
and that is why they can be considered separately.
We use the parameterization
\be
G\left(\hat n,\hat \varphi\right)
=\int_0^{2\pi} \frac{d\theta}{2\pi}
e^{-i\theta\hat n}
\tilde G\left(\theta,\hat \varphi\right).
\mylabel{Gangle}
\ee
Commutation relation \rref{Com2} gives
\[
e^{ix\sin\hat\varphi}e^{-i\theta \hat n }
= e^{-i\theta \hat n }e^{ix\sin(\hat\varphi+\theta)}.
\]
Thus, using definitions of \req{LSCBA1} we obtain
\be
\begin{split}
&\hat U_\perp \hat G\left(\hat n; \hat\varphi; {\hat\R}(\q)\right)
\hat U_\perp^\dagger
=\int_0^{2\pi} \frac{d\theta}{2\pi}
e^{-i\theta\hat n}
\tilde G\left(\theta,\hat \varphi\right)
\\
&
\quad \times
\exp\Bigg\{
iqR_c\left[\sin(\hat\varphi_q+\theta)
-\sin\hat\varphi_q
\right]
\\
&
\quad
-\frac{iq^2R_c
[\sin(2\hat\varphi_q+2\theta)-\sin2\hat\varphi_q]
}{4p_F}
\Bigg\}
\end{split}
\mylabel{perp1}
\ee
The characteristic values of $q$ entering into the integral
can be estimated as $qR_c\simeq R_c/\xi \gg R_c/\lH =\sqrt{2N} \gg 1$.
Let us call the argument of the exponent $\beta(\theta,\phi)$.
Because, $qR_c \gg 1$, the integrals will be determined by
 the saddle point determined by
$\beta^\prime_\theta(\theta,\phi)=0$, and $\beta^\prime_\phi(\theta,\phi)=0$
the latter condition gives $\theta=0$.
Thus, we write
\[
e^{i \beta(\theta,\phi) } \approx
{2\pi}\delta(\theta)\delta\left[\beta^\prime_\theta(\theta,\phi)\right],
\]
in a sense that the saddle point integration on the LHS
gives
the same result as the integration in the RHS.
Employing this approximation in \req{perp1} and taking into account
$\parallel \hat n \parallel \ll qR_c$,
\be
\hat U_\perp \hat G\left(\hat n; \hat\varphi\right)
\hat U_\perp^\dagger
=\frac{\tilde G\left(0,\hat \varphi\right)}{{qR_c}}
\delta\left(\cos\hat\varphi_q -\frac{q\cos
2\hat\varphi
_q}{2p_F}\right)
,
\mylabel{perp2}
\ee
where $\delta-$function has to be understood in an operator sense.

We substitute \req{perp2} into \req{LSCBA1} and use \req{Gangle}
to find $\tilde G(\theta=0)$. With the help of the
commutation relation \rref{Com2} and using the small
angle scattering condition \rref{smallangle}, we obtain
\be
\begin{split}
 \hat\Sigma
 \left(\hat\varphi;{\hat\R}\right)
 &=\frac{-i}{2\pi \v_F}
\int_{-\infty}^\infty
\frac{d q}{2\pi}
W(q)
\exp
\left[-iq
 \bzeta_{12}\cdot \hat{\ep}
\nn_q\right]
\\
&
\times
 \hat{g}\left[ \hat\varphi
+\frac{q}{p_F}
; {\hat\R}-\lH^2 q\nn_q\right]
\\
\nn_q&\equiv \left[\cos
\left(\hat\varphi+\frac{q}{2p_F}\right),\,
\sin\left(\hat\varphi+\frac{q}{2p_F}\right)\right],
\end{split}
\mylabel{LSCBA3}
\ee
where we introduced the analog of the Green function
in the Eilenberger equation
\be
\hat g \left( \hat\varphi
; {\hat\R}\right)
=i\w_c
\sum_k\hat G \left(\hat{n}+k; \hat\varphi
; {\hat\R}\right).
\mylabel{smallg}
\ee
Notice, that operator $\hat \varphi$ commutes with all other
operators entering into \req{LSCBA3}, and, therefore, it can be
treated as a $c$-number.

Equation \rref{LSCBA3} also can be rewritten in a different form
\be
\hat\Sigma_{t,t'}=-i\hat {\cal K}_{t,t'}\hat g(t,t';\varphi,{\hat \R}),
\mylabel{newform}
\ee
where the kernel ${\cal K}_{t,t'}$ is
\be
{\cal K}_{t,t'}=\int  \frac{W(q) }{4\pi^2 v_F}
\exp\left[\frac{q\partial_\varphi}{2p_F}\right]
e^{-iq \P \nn(\varphi)}
\exp\left[\frac{q\partial_\varphi}{2p_F}\right]
dq,
\label{kernelK}
\ee
with
\be
\P=\bzeta_{t,t'}\hat\varepsilon-i\lambda_H^2{\nnabla_{\R}}
\label{Pvector}
\ee
This form is more convenient for the further expansion in $1/\xi p_F$, as
we show later.

Closing this subsection we write down the expressions
for the variation of the electron density, $\delta N_e$,
 and the current density, $\j$.
Representing the coordinate operators by Eq.~(\ref{H}) and
approximating the  operators with the
help of \req{parexpand} we obtain
\begin{subequations}
\label{rhoj}
\be
\begin{split}
&\delta N_e\left(\r,t\right)=-m_e\int\limits_0^{2\pi}
\frac{d\varphi}{2\pi}
\delta g^K\left[t,t;\varphi,\r_g\right];
\\
&
\r_g=\r-R_c\hat\ep\nn(\varphi)-\bzeta(t),
\end{split}
\mylabel{rho}
\ee
and $\delta g^K$ denotes the deviation of the Keldysh Green function
from its equilibrium value in the absence of the external fields.
The coordinate $\bzeta(t)$ is defined in \req{zeta}.

For the full electric current we have
\be
\begin{split}
& \j(\r,t) = eN_e\left(\r,t\right) \pt \bzeta+ \j^{(d)}(\r,t),
\end{split}
\ee
where the first term in the current is  dissipationless.
At constant electric field it is nothing but the Hall current.
The second term is given by
\be
\begin{split}
&\j^{(d)}(\r,t)=- e p_F\int\limits_0^{2\pi}
\frac{d\varphi}{2\pi}\nn(\varphi)
g^K\left[t,t;\varphi;\r_g\right].
\end{split}
\mylabel{j}
\ee
\end{subequations}
The numerical coefficients
in \reqs{rho} and \rref{j} are written with account of the
spin degeneracy, and $N_e$ is the total electron density.

\subsection{Equation for the spectrum}

In this subsection we solve equation
\rref{Gr3} with the Hamiltonian \rref{H2} and
the semiclassical self-energy \rref{LSCBA3}
\be
\begin{split}
&\left\{i\pt - \w_c\left[\hat n + \delta(\mu)\right]
\right\}
 G^R\left(t,t_1; \hat{n}\right)
=
\frac{\delta(t-t_1)}{2\pi}
\\
&
+\int_{t_1}^{t}dt_2\Sigma^R\left(t,t_2\right)
G^R\left(t_2,t_1; \hat{n}\right).
\end{split}
\mylabel{trgr1}
\ee
Hereinafter, we  suppress $\varphi$ and $\R$ arguments in the Green
function and the self-energy whenever they are the same
in the both sides of the equations.
Our purpose is to represent the $G^R$ in terms of the Green
functions \rref{smallg} only.

For the calculation of the spectrum it is sufficient to
keep the terms only to the zeroth order in small parameter $q/p_F$.
Equations \rref{LSCBA3} and \rref{newform} then simplify to

\be
\begin{split}
 i\hat\Sigma^R
 \left(\hat\varphi;{\hat\R}\right)
 &=
h_1\left[
\frac{
\left(\bzeta_{12}\hat\ep
-i\lH^2\nnabla_R\right)\cdot \nn(\varphi)}{\xi}
\right]
\frac{ \hat{g}^R\left[ \hat\varphi
; {\hat\R}\right]}{\tau_q}.
\end{split}
\mylabel{sigmaR}
\ee
Here
\be
\frac{1}{\tau_q}=
\frac{1}{2\pi v_F}\int\frac{dq}{2\pi}W(q);
\mylabel{tauqa}
\ee
is nothing but the standard Born approximation for the
quantum scattering time for small angle scatterers,
and the dimensionless function
\be
h_1\left(x\right)=
\frac{\int{dq}W(q)e^{iq x\xi}}{\int{dq}W(q)} = \frac{1}{1+x^2}
\mylabel{h1}
\ee
characterizes the effect of the external electric field during the
cyclotron motion between electron returns to the same impurity.

We will look for the self-energy in the form
\be
i\Sigma^R(t,t_1)=
\frac{\delta\left(t-t_1\right)}{2\tau_q} +\sum_{l=1}^\infty\lambda^l
{\cal S}^R_l\left(\hat{\cal T}^lt\right)
\delta\left(\hat{\cal T}^lt-t_1\right),
\mylabel{smallsigma}
\ee
where the coherence factor $\lambda$
describes
 the phase accumulation during one
period and it is defined as
\be
\lambda=\exp\left(-\frac{\pi }{\w_c\tau_q}-i2\pi \delta(\mu)\right),
\mylabel{lambda}
\ee
and ${\cal S}^R_l$ are to be found self-consistently.
The time shift operator is defined as
\be
\hat{\cal T}^l t=t-\frac{2\pi l}{\w_c}.
\mylabel{tshift}
\ee

We look for the
solution of \req{trgr1} in the form
\be
\begin{split}
& i G^R\left(t,t_1; \hat{n},\hat\varphi\right)=
\frac{e^{-i\w_c\delta(\mu)\left(t-t_1\right)}
e^{-\frac{t-t_1}{2\tau_q}} }{2\pi}
\\
&
\quad \times
\sum_l
\Bigg\{\left[e^{-i\w_c\hat n \, t}
{\cal G}^R_l
\left(\hat{\cal T}^lt,t_1,\hat\varphi\right)
e^{i\w_c\hat n \, t_1}\right]
\\
&
\quad
\times
\theta\left(\hat{\cal T}^lt-t_1\right)
\theta\left(t_1-\hat{\cal T}^{l+1}t\right)
\Bigg\}.
\end{split}
\mylabel{ansatz}
\ee

Substituting \reqs{ansatz}  and \rref{smallsigma} into \req{trgr1},
and using $e^{i2\pi l\hat n}=1$, $e^{ix\hat n}\varphi e^{-ix\hat n}
=\varphi + x $
we obtain the chain of equations
\be
\begin{split}
&{\cal G}^R_l\left(t,t_1; \hat\varphi\right)
=g^R_{l}\left(t_1; \hat\varphi\right)
\\
&
 -\int\limits_{t_1}^{t}dt_2\sum_{m=1}^l{\cal S}^R_{m}
\left[
{\cal T}^{m-l}
t_2;\hat\varphi+\w_c\left(t_2-t_1\right) \right]
{\cal G}^R_{l-m}\left(t_2,t_1; \hat\varphi \right)
\end{split}
\mylabel{trgr2}
\ee
where
\be
\begin{split}
&
g^R_{l}\left(t; \hat\varphi, {\hat\R}\right) \equiv
{\cal G}^R_l\left(t,t; \hat\varphi, {\hat\R}\right);
\\
&
{\cal G}^R_l\left(t,t_1\right)={\cal G}^R_{l-1}\left(\hat{\cal T}t,t_1\right);
\end{split}
\mylabel{gsmall2}
\ee
with the initial condition ${g}_0^R=1$.

Final step is the self-consistency procedure which amounts into
substitution of \req{ansatz} into \req{smallg}.
It gives
\footnote{
In the derivation of \req{gsmall3}, the uncertainty $\theta(x)\delta(x)$
had to be resolved as $\delta(x)/2$.}
:
\be
\begin{split}
& g^R \left(t;t_1\right)=
\frac{\delta(t-t_1)}{2}
 +
\sum_{l=1}^{\infty}\lambda^l
g^R_{l}\left(t_1\right)
\delta\left(\hat{\cal T}^lt-t_1\right)
,
\end{split}
\mylabel{gsmall3}
\ee
where $\lambda$ is defined in \req{lambda}.
Using \req{gsmall3} in \req{sigmaR} and extracting ${\cal S}_l$
of \req{smallsigma}, we find
\be
{\cal S}_l^R\left(t,\varphi \right)
=
h_1\left[
\frac{
\left(\bzeta_l(t)\hat\ep
-i\lH^2\nnabla_R\right)\cdot \nn(\varphi)}{\xi}
\right]
\frac{g^R_{l}\left(t,\varphi\right)}{\tau_q}.
\mylabel{smallsigma4}
\ee
We use the short hand notation
\be
\bzeta_l(t)\equiv
\bzeta\left(\hat{\cal T}^{-l}t\right)
- \bzeta\left(t\right),
\mylabel{zetal}
\ee
where the finite time shift operator is defined in \req{tshift}.

Equations \rref{trgr2} -- \rref{smallsigma4} constitute a complete system for
the  spectrum averaged over disorder.
Note, that the Green function ${\cal G}_l$ is
expressed in terms of  ${\cal G}_{m}$ with $m<l$ only.

\subsection{Equation for the distribution function}
\label{sec:df}

In this subsection we will reduce \reqs{keq1} to the canonical
Boltzmann form. According to \req{LSCBA3} the self-energies
$\hat \Sigma$ do not depend on $\hat{n}$. It suggests that the distribution
function $\hat f$ does not depend on $\hat n$ either.
This observation enables us to substitute \req{Gr5} into \req{smallg}
and perform the summation over $k$
with the help of \req{gsmall3} and relations
 $g^A=-\left[g^R\right]^\dagger, \ f=f^\dagger$.
We find
\begin{widetext}
\be
\begin{split}
\frac{1}{2}
 \left[g^R-g^A
-g^K\right]_{t,t'}=
f\left(t,t'; \varphi, \R \right)
+ & \sum_{l=1}^{\infty}\lambda^l
g^R_{l}\left(\hat{\cal T}^l t; \varphi, \R \right)
f\left(\hat{\cal T}^lt,t',\varphi, \R\right)\\
& \quad
+ \sum_{l=1}^{\infty}\left(\lambda^l\right)^*
f\left(t,\hat{\cal T}^lt',\varphi, \R\right)
\left[g^R_{l}\left(\hat{\cal T}^l t';
\varphi, \R \right)\right]^\dagger
,
\end{split}
\mylabel{gksmall}
\ee
where the coherence factor $\lambda$ is defined in \req{lambda},
and the time shift operator is given by \req{tshift}.

Substitution of \req{gksmall} into \reqs{rho} and \rref{j}
yields the connection
between the distribution function $f$ and the observables:
\begin{subequations}
\label{rhoj1}
\be
\begin{split}
&\delta N_e\left(\r,t\right)=2 m_e\int\limits_0^{2\pi}
\frac{d\varphi}{2\pi}
\Bigg\{\delta f\left(t,t; \varphi,\r_g
\right)
+  2\Re \sum_{l=1}^{\infty}\lambda^l
g^R_{l}\left(\hat{\cal T}^lt; \varphi, \r_g \right)
 f\left(\hat{\cal T}^lt,t; \varphi,\r_g\right)
\Bigg\};
\end{split}
\mylabel{rho1}
\ee
\be
\begin{split}
&\j^{(d)}(\r,t)=2 e p_F\int\limits_0^{2\pi} \frac{d\varphi}{2\pi}\nn(\varphi)
\Bigg\{f\left(t,t; \varphi,\r_g
\right)
+  2\Re \sum_{l=1}^{\infty}\lambda^l
g^R_{l}\left(\hat{\cal T}^lt; \varphi, \r_g \right)
 f\left(\hat{\cal T}^lt,t; \varphi,\r_g\right)
\Bigg\}
,
\end{split}
\mylabel{j1}
\ee
where we used the short hand notation
$
\r_g=\r-R_c\hat\ep\nn(\varphi)-\bzeta(t).
$
\end{subequations}

To derive the kinetic equation, we substitute $f$ into
\reqs{keq1}. Equation \rref{Gr10} gives
\begin{subequations}
\label{kineq}
\be
\left[\frac{\partial}{\partial t}
+ \frac{\partial}{\partial t'}
+ \w_c \frac{\partial}{\partial \varphi}
\right]f\left(t,t'; \varphi,\R\right)
= \hat\St\{\,f\}_{t,t'}.
\mylabel{mf1}
\ee
According to \req{Gr10a}, the collision integral is defined in
terms of the electron self-energy, the latter is given by
Eq.~(\ref{newform}). Substituting Eq.~(\ref{newform}) into
\req{Gr10a} and using the relation \req{gksmall}, we obtain the
following expression for the collision integral:
\be
\begin{split}
 \St\{f\}_{t,t'}
=&
-\left\{ \hat {\cal K}(t,t')
 f\left(t,t'\right)\right\}
\\
&
+
\sum_{l=1}^\infty
\lambda^l \Bigg\{
\left[\hat{\cal K}(t,\hat{\cal T}^lt)g^R_l\left(\hat{\cal T}^lt\right)
\right]
f\left(\hat {\cal T}^lt,t'\right)
 -\left[\hat {\cal K}(t,t')f\left(\hat {\cal T}^lt,t'\right)
 g^R_l\left(\hat{\cal T}^lt\right)
\right]
 \Bigg\}
\\
&
-
\sum_{l=1}^\infty
(\lambda^*)^l \Bigg\{
f\left(t,\hat {\cal T}^lt'\right)
\left[\hat{\cal K}(\hat{\cal T}^lt',t')
g^A_l\left(\hat{\cal T}^lt'\right)
\right]
 -\left[\hat {\cal K}(t,t')f\left(t,\hat {\cal T}^lt'\right)
 g^A_l\left(\hat{\cal T}^lt'\right)\right]\Bigg\},
\end{split}
\label{STgeneral}
\ee
\end{subequations}
\end{widetext}
In \req{STgeneral} kernel $\hat{\cal K}$
is given by \req{kernelK}, functions $g_l^{R,A}$ are defined in the
previous subsection by \req{trgr3},
the coherence factor $\lambda$ is defined in \req{lambda},
and the time shift operator is defined in \req{tshift}.
Note that we suppressed $\varphi$ and $\R$ arguments in the entries
of \reqs{STgeneral} for brevity.
The first line in \req{STgeneral} corresponds to the classical
scattering off an impurity; the second and the third lines
describe the retarded interference corrections due to
the returning orbits.

One property of the kinetic equation~\rref{kineq} is worth
emphasizing because
it is a crucial check of the consistency
of the approximation we made. Consider the constant electric field $\E$, so
that
\be
\bzeta(t)=-c t\frac{\hat \ep \E}{B}.
\label{dcfieldzeta}
\ee
Then the distribution function
\[
f(t,t')=f_F(t-t')e^{-i e\E\hat{\R}(t-t')}
\]
where the equilibrium distribution function is given by \req{Gr5a},
null both the collision integral and the left hand side of
\req{mf1}. In the energy representation
$f(\ep)=f_F(\ep+e\E\hat\R)$ corresponds to the thermodynamic
equilibrium of the system in the moving coordinate frame.

For the small angle scattering, see \req{smallangle}, we
expand the rotation operator $\exp(q\partial_\varphi/2p_f)$
in  the powers of $q/p_F$ and obtain the following expression for
the kernel \req{kernelK}:
\be
\hat {\cal K}=\hat {\cal K}_\bot+
\hat {\cal K}_j+\hat {\cal K}_\varphi+\hat k_O
\label{Kexpansion}
\ee
where the first term contains only even angular harmonics:
\begin{subequations}
\label{St}
\be
\begin{split}
\hat {\cal K}_\bot = &
\frac{\left([p_F\bzeta_{t,t'}\hat\ep -
i R_c\nnabla_R]\cdot \nn \right)^2}{\tau_{tr}}
\\
& \times
h_1\left(
\frac{\left(\bzeta_{t,t'}\hat\ep
-i\lH^2\nnabla_R\right)\cdot \nn}{\xi}
\right),
\label{mf7}
\end{split}
\ee
the second term contains odd angular harmonics:
\be
\begin{split}
\hat {\cal K}_j & =  \frac{
\left[R_c\nnabla_R \hat\ep
-i p_F\bzeta_{t,t'}
\right]\cdot\nn
}
{\tau_{tr}}h_2\left(
\frac{
\left(\bzeta_{t,t'}\hat\ep
-i\lH^2\nnabla_R\right)\cdot\nn}{\xi}
\right)
\\
& +
\frac{p_F\xi}{\tau_{tr}}
h_3\left(
\frac{\left(\bzeta_{t,t'}\hat\ep
-i\lH^2\nnabla_R\right)\cdot \nn}{\xi}
\right)\partial_\varphi ,
\label{mf8}
\end{split}
\ee
and the third term represents the angular diffusion:
\be
\begin{split}
&\hat {\cal K}_\varphi =-
\frac{1}{\tau_{tr}}
\partial_\varphi h_2\left(
\frac{\left(\bzeta_{t,t'}\hat\ep
-i\lH^2\nnabla_R\right)\cdot \nn}{\xi}
\right) \partial_\varphi
.
\label{mf5}
\end{split}
\ee
The remaining term $\hat k_{O}$ describes contributions
which are of the order of
$1/(p_F\xi)^2$ smaller.
\end{subequations}
In \reqs{St}  we introduced the transport mean free time
\be
\frac{1}{\tau_{tr}}=
\frac{1}{4\pi v_Fp_F^2}\int\frac{dq}{2\pi}q^2W(q)
\simeq \frac{1}{\tau_q}\frac{1}{(p_F\xi)^2},
\mylabel{tautr}
\ee
and the dimensionless function
\be
 h_2\left(x\right)=
 \frac{\int{dq}q^2W(q)e^{iq x\xi}}{\int{dq}q^2W(q)} =
 \frac{1-3x^2}{\left(1+x^2\right)^3}.
 \mylabel{h2}
\ee
Function $h_1(x)$ is defined by \req{h1}, and
$\nn = \left(\cos\varphi, \sin\varphi\right)$.

Let us discuss in more details the meaning of
the components of the kernel \req{Kexpansion}.
The third term in \req{Kexpansion}, given by \req{mf5}, describes
the angular diffusion. Its contribution to the collision integral
suppresses angular harmonics of the distribution function rather
than the zeroth harmonics.

The second term $\hat {\cal K}_j$ in \req{Kexpansion}, see \req{mf8},
represents the scattering
process accompanied by simultaneous creation of odd-angular
harmonics and energy shift. It is this term that is responsible
for the dissipative current. When its is substituted into the
collision integral \req{STgeneral}, the first line of the collision
integral describes an instantaneous scattering and gives the
classical conductivity. The second and third
lines describe the interference due to the returns of the
cyclotron trajectories.
All non-linear effects considered
in the following sections originate from the second and third lines
of the collision integral \req{STgeneral} with the full kernel $\hat {\cal K}$
replaced by $\hat {\cal K}_j$.
The linear in $\partial_\varphi$ term in \req{mf8}
may be safely omitted
since it does not give rise to any effects
relevant for the future consideration. We kept it in \req{mf8}
to display explicitly that the operator $\hat {\cal K}_j$ is Hermitian
as guaranteed by the relation
\[
\frac{d h_3(x)}{d x}=2ix h_2(x).
\]

Finally, the first term $\hat {\cal K}_\bot$
of the kernel \req{Kexpansion}, see \req{mf7},
is responsible for the evolution of the distribution function perpendicular
to the Fermi surface, which does not mix angular harmonics
of the distribution function with different parity.
Similarly to $\hat {\cal K}_j$, in the first line of the collision integral
\req{STgeneral}, $\hat {\cal K}_\bot$
describes the classical effect of the electric field
on the electron distribution -- Joule heating.
The other terms in \req{STgeneral} describe the
returns of the cyclotron trajectories,
which may result in oscillating components of the distribution
function with period $\hbar\w_c$.\cite{Mirlin-ac,inelastic}
We remark that term \req{mf7} of the collision integral
can not be considered separately from inelastic
processes such as the electron-electron  and electron-phonon
interactions. Indeed, taking the zeroth angular harmonics of \reqs{mf1}
and \rref{mf7}, one finds the correction to the distribution
function, that grows infinitely in time. This is just a signature
of the energy absorbed by the system from the external field.
Elastic impurities alone can not stabilize the distribution function.
The electron-electron interaction suppresses  large deviations
from the Fermi distribution with some effective $T_{eff}$ whereas the
electron-phonon interaction prevents  $T_{eff}$ from an infinite increase.

In the remainder of the paper
we will consider only the phenomena associated with $\hat {\cal K}_j$,
that are not sensitive to effects  of
the external field on the distribution function. Therefore
we neglect the contribution to the collision integral \req{STgeneral}
originating from the term Eq.~(\ref{mf7}) of the
kernel \req{Kexpansion}. This contribution may be neglected if the
energy relaxation time is small, the condition which is generally
not valid.
In recent preprint\cite{Mirlin-ac}
an estimate of the contribution to the dc resistivity from
oscillating component of the distribution function
was presented. When inelastic processes are weak, this
contribution is larger than the contributions studied in the
present paper. Nevertheless, the effects considered here are
robust and are described by different system parameters,
therefore they deserve a separate consideration.
The contribution which depends
on the form of the electron distribution function will be
presented elsewhere.\cite{inelastic}
The implicit assumption everywhere
will also be $(T,T_{eff}) \ll \mu$, i.e. the electron system is degenerate.

\section{Linear transport}
\label{Linear}

The purpose of this section is two-fold:
(i) to demonstrate how the solution of the QBE is obtained
for the simplest case and to reproduce relatively known
results; (ii) to derive formulas for the spectrum which
can be used as  building blocks for consideration of
more elaborate effects in the further sections.

We begin with the solution of \req{trgr2} for the spectrum.
In the linear regime, we can put $h_1=1$ in \req{smallsigma4}.
After this simplifications entries of \reqs{trgr2} become independent of
the angle $\varphi$ and time $t_1$. We find with the help of \reqs{gsmall2}
and \rref{smallsigma4}
\be
\begin{split}
&{\cal G}^R_l\left(t\right)
=g^R_{l}
 -\frac{1}{\tau_q}\sum_{m=1}^l\int_{0}^{t}dt_1 g^R_{m}
{\cal G}^R_{l-m}\left(t_1 \right),
\\
&
g^R_{l} =
{\cal G}^R_l\left(0\right)={\cal G}^R_{l-1}\left(\frac{2\pi}{\w_c}\right);
\end{split}
\mylabel{trgr3}
\ee
with the initial conditions $g^R_{0}=1, \ g^R_{-1}=0$.

Non-linear recursion relations \rref{trgr3}
can be resolved exactly with the result
\be
\begin{split}
&{\cal G}_l(x)=\frac{L_{l}^1\left[(x+l)\alpha\right]}{l+1}
+\frac{\alpha (1-x) L_{l-1}^2\left[(x+l)\alpha\right]}{l+1};
\\
&
g_l=\frac{L_{l-1}^1\left(l\alpha\right)}{l};
\quad x=\frac{\w_c t}{2\pi};
\quad
\alpha=\frac{2\pi}{\w_c\tau_q},
\end{split}
\mylabel{sol1}
\ee
where $L_l^m(x)$ is the Laguerre polynomial \cite{mathbook}.

Next step is to find the distribution function. To do so
we use \req{Gr5a} for equilibrium distribution and
solve \req{mf1} to the leading order in  $1/(\w_c\tau_{tr}) \ll 1$
and in the first order in $\bzeta (t)$, see \req{dcfieldzeta}.
In this approximation
only the collision term \rref{mf8} contributes.
Taking the limits
$t_2 \to t_1 -\frac{2\pi l}{\w_c}$ for integer $l$, we find
\be
\begin{split}
2\pi \delta& f\left({\cal T}^lt,t; \varphi\right)
=
\frac{-\pt +\w_c\partial_\varphi}
{\tau_{tr}\left(\pt^2+\w_c^2\right)}
\Bigg\{
\left[p_F\pt\bzeta(t) \cdot \nn(\varphi) \right]
\lambda_l^* g^R_l
\\
&
+
\sum_{m=0,\ m\neq l}^\infty
\frac{
\left[ p_F\bzeta_{l-m}(t) \cdot \nn(\varphi) \right]
\left(\lambda^m\right)^* \pi T}
{\sinh\frac{2\pi^2 T(l-m)}{\w_c}}
g^R_m
\\
&
+
 \sum_{m=1}^\infty
\frac{
\left[p_F\bzeta_{l+m}(t)\cdot\nn(\varphi) \right]
\lambda^m \pi T}
{\sinh\frac{2\pi^2 T(l+m)}{\w_c}}
 g^R_m
\Bigg\}
,
\end{split}
\mylabel{solf}
\ee
where $\bzeta_l(t)$ is defined in \req{zetal}.

Equations \rref{sol1} and \rref{solf}
are sufficient to calculate the linear response of the electric
current \req{j1} to the applied electric field within the
self consistent Born approximation at
arbitrary temperatures and magnetic fields. First we discuss the
high temperature limit, $T \gg \hbar\w_c$, when the conductance
is a smooth function of the applied magnetic field.
Then we take into
account $1/\omega_c$ oscillations of the conductivity
which appear at $T/\w_c\gg 1$
(Shubnikov--de Haas oscillations).

\subsection{$Ac$- and $dc$- transport at high temperatures}
\label{sec:acdc}

At $T\gg\w_c$ only the first term in \req{solf} remains and
all other terms are exponentially suppressed.
Introducing $N_e=p_F^2/(2\pi)$ and  substituting
\reqs{sol1} and \rref{solf} into \req{j1}, we find with the help of
\req{lambda}
\be
\begin{split}
&\j^{(d)}(t)
=
\frac{\left(\hat{\varepsilon}\pt + \w_c \right)^2}
{(\pt^2+\w_c^2)^2}
\left(\frac{e^2N_e}{m_e\tau_{tr}}\right)
\\
&\times
\Bigg\{
\E(t)
+2
\sum_{l=1}^\infty
 \frac{e^{-\frac{2\pi l}{\w_c\tau_q}}}{l^2}
\left[L_{l-1}^1\left(
\frac{2\pi l}{\w_c\tau_q}
\right)\right]^2
\E\left(\hat{\cal T}^lt\right)
\Bigg\},
\end{split}
\mylabel{Lin1}
\ee
where the time finite shift operator is given by \req{tshift}.
First term in \rref{Lin1} describes usual scattering contribution
and the subsequent terms result from the multiple returns to the
same impurity.
In the frequency representation \req{Lin1} may be written as
\[
\j_\w^{(d)} = \hat\sigma^{(d)}(\w)\E_\w,
\]
where $\hat\sigma^{(d)}(\w)$ is the conductivity tensor:
\be
\begin{split}
&\hat\sigma^{(d)}(\w)
=
\frac{\left(-i\w\hat{\varepsilon} + \w_c \right)^2}
{(\w_c^2-\w^2)^2}
\left(\frac{e^2N_e}{m_e\tau_{tr}}\right)
\\
&\times
\Bigg\{
1
+2
\sum_{l=1}^\infty
\frac{
 e^{-\frac{2\pi l}{\w_c\tau_q}+i\frac{2\pi\w l}{\w_c}}}{l^2}
\left[{L_{l-1}^1\left(
\frac{2\pi l}{\w_c\tau_q}
\right)}\right]^2\Bigg\}.
\end{split}
\mylabel{Lin1a}
\ee

\begin{figure}
\epsfxsize=0.45\textwidth
\epsfbox{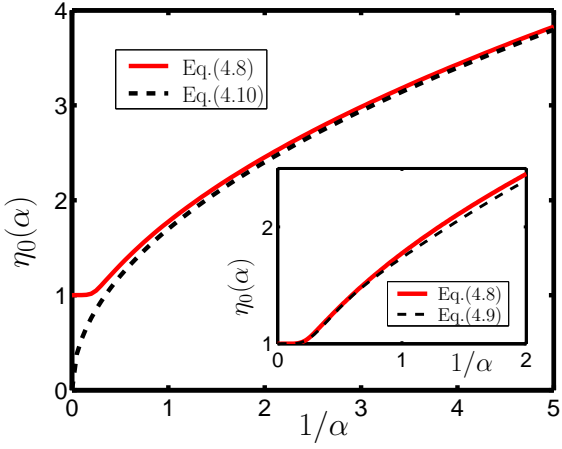}
\caption{
The solid line represents the high temperature
magnetoresistance curve \req{Lin8}, $\alpha=2\pi/(\w_c\tau_q)$.
The dashed curve in the main panel
is the high field approximation \rref{Lin8b}.
The dashed curve in the inset is the
low-field asymptotic expression  \rref{Lin8a}.
}
\label{fig:Lin8}
\end{figure}

At $\omega=0$, \req{Lin1a} gives the non-oscillating correction
to the diagonal $dc$ resistivity $\rho_{xx}$.
Using $\w_c\tau_{tr} \gg 1$,
one writes $\rho_{xx}=\sigma_{xx}(\w=0)\left[\rho_{xy}\right]^2$ and
\req{Lin1a} yields
\be
\mylabel{Lin8}
\begin{split}
\rho_{xx}(B)=\rho_{D} \eta_0\left(\frac{2\pi}{\w_c\tau_q}\right),
\end{split}
\ee
where
\begin{equation}
\rho_{D}=\frac{m_e}{e^2N_e\tau_{tr}}
\mylabel{rhoD}
\end{equation}
is the Drude resistivity, and
\begin{equation}
\eta_0(\alpha)=
1+2\sum_{l=1}^\infty
\frac{e^{-\alpha l}}{l^2}
\left[{L_{l-1}^1\left(
\alpha
\right)}\right]^2.
\mylabel{eta0}
\end{equation}
The asymptotic behavior of \req{eta0} is
\be
\eta_0(\alpha) =
1 + 2
\left[e^{-{\alpha}}+
 e^{-2\alpha}
\left(1-\alpha
\right)^2\right]
+{\cal O}\left(e^{-3\alpha}\right)
\mylabel{Lin8a}
\ee
at $\alpha \gg 1$ (weak magnetic field),
and
\be
\eta_0(\alpha)=
\frac{16}{3\pi}
\sqrt{\frac{1}{\alpha}}
+{\cal O}\left(\sqrt{\alpha}
\right),
\mylabel{Lin8b}
\ee
at $\alpha \ll 1$ (strong magnetic field).
Function $\eta_0(\alpha)$ together with
its asymptotes is plotted in Fig.~\ref{fig:Lin8}.

Equation \rref{Lin8} was written to the leading order in
$1/(\w_c\tau_{tr})$. To obtain the correction to the Hall
coefficient one has to solve \req{mf1} and
take into account the term \rref{mf5} in the first order perturbation
theory. The Hall coefficient can be expressed in terms
of the third powers of the coefficients (\ref{sol1}). The final result,
however, will have the smallness $1/(\w_c\tau_{tr})^2$ and that
is why we will not write down the explicit form of those corrections.
The Hall coefficient in this approximation is
 \be
 \rho_{xy}=\frac{B}{ecN_e }
\left[1+\frac{{\cal F}_H(\w_c\tau_q)}{\w_c^2\tau_{tr}^2}\right]
 \mylabel{Hall}
 \ee
where ${\cal F}_H(x)$ vanishes exponentially at $x\to 0$.

At finite frequency $\w$, \req{Lin1} gives, in particular, the oscillatory
dependence of the  absorption of microwave radiation with field
$\E(t)=Re \E_\w e^{-i\w t}$ on frequency
$\w$.
We find
\be
\begin{split}
&\Re\ \E^*_\w\sigma_{xx}(\w)\E_\w
= \left(\frac{e^2N_e |E_\w|^2}{m_e\tau_{tr}}\right)
\\
&
\times
\frac{\w_c^2+\w^2-2\w\w_c\cos 2\beta}
{(\w_c^2-\w^2)^2}
{\cal F}_1\left(\frac{\w}{\w_c},\frac{2\pi}{\w_c\tau_q}\right),
\end{split}
\mylabel{Lin9}
\ee
where
parameter $\beta$ describes
the polarization of the field by parameterization
of $\E_\w$ as
\be
\begin{split}
&\frac{\sqrt{2}\E_\w}{ \sqrt{\E_\w\cdot\E_\w^*}}
=\mm_+\cos\beta+\mm_-\sin\beta
,
\ \mm_\pm =\mm \pm i\hat\ep\mm,
\end{split}
\mylabel{wf9}
\ee
where $\mm$ is the unit vector.
For the circular polarization of the microwave
$\beta=0, \pm\pi/2$. For the linear polarization
along $\mm$, $\beta=\pi/4$.

The dimensionless function ${\cal F}_1(w,\alpha)={\cal F}_1(w+1,\alpha) $
represents the normalized coefficient of microwave absorption:
\be
{\cal F}_1\left(w,\alpha\right)=
1+2\sum_{l=1}^\infty
\frac{e^{- l \alpha }\cos 2\pi l w}{l^2}
\left[L_{l-1}^1\left(l \alpha
\right)\right]^2.
\mylabel{Lin9a}
\ee
At weak field, $\alpha\gg 1$, function ${\cal F}_1(w,\alpha)$
is well described by the first few terms:
\be
{\cal F}_1\left(w,\alpha \right)\approx
1+2
e^{-  \alpha}\cos 2\pi w+2
e^{-  2 \alpha}(1-\alpha)^2\cos 4\pi w
+\dots
.
\mylabel{Lin9b}
\ee
At strong magnetic field, $\alpha\ll 1$, we use the asymptotic
expression of the Laguerre polynomials in terms of the Bessel
functions $J_n(y)$\cite{mathbook} and obtain
\be
\frac{e^{- l x/2}}{l}
L_{l-1}^1\left(l x
\right)\approx \frac{J_1\left(2l\sqrt{x}\right)}{l\sqrt{x}}, \quad
l \ll \frac{1}{x}.
\mylabel{gapprox}
\ee
Substituting \req{gapprox} in \req{Lin9b} and employing the Poisson summation
formula we obtain for $\alpha\ll 1$
\be
{\cal F}_1\left(w ,\alpha\right)=\frac{16}{3\pi\sqrt{\alpha}}
\sum_{k=-\infty}^\infty  {\cal H}_1
\left(\frac{\pi \left|w-k\right|}{\sqrt{\alpha}}\right)
\mylabel{Lin9c}
\ee
where
\[
\begin{split}
{\cal H}&_1(x)=
\frac{3\pi\theta\left(
2-|x|\right)}{4}\int_0^\infty
d y\cos x y \left[\frac{J_1(x)}{x}\right]^2
\\
&=
\frac{\left(2+x\right)\theta\left(
2-|x|\right) }{8}
\\
&\times
\Bigg\{\left(4+x^2\right)
E\left(\frac{2-x}{2+x}\right) - 4x K\left(\frac{2-x}{2+x}\right)
\Bigg\}.
\end{split}
\]
Functions $K(x)$ and $E(x)$ are complete elliptic integrals of the
first and  second kind respectively, and
function ${\cal H}_1(x)$ can also be obtained as a convolution
of two semicircle densities of states.
Equation  \rref{Lin9} and its asymptotes
\rref{Lin9b} and \rref{Lin9c} are consistent at $|\w-\w_c|\tau_{tr}\gg 1$
with the result of recent preprint.\cite{Mirlin-ac}

\begin{figure}[h]
\epsfxsize=0.45\textwidth
\epsfbox{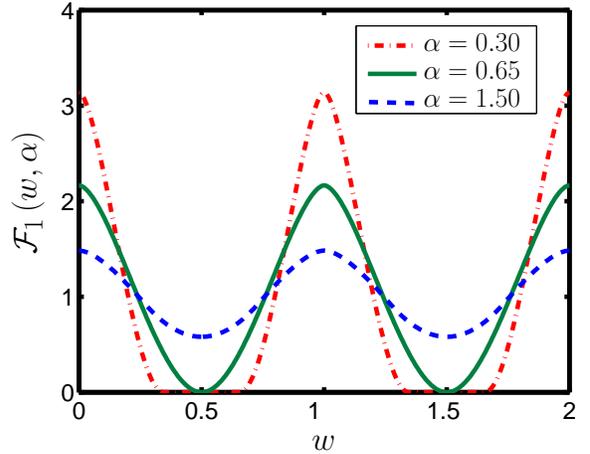}
\caption{
Dependence of normalized microwave
absorption  coefficient ${\cal F}_1(w,\alpha)$, see \req{Lin9a},
on frequency $w=\w/\w_c$ at three values of
$\alpha=\frac{2\pi}{\w_c\tau_q}$.
The function ${\cal F}_1(w,\alpha)$ is periodic in $w$
with the period $1$. Note, that at strong magnetic
field ($\alpha=0.3$) function vanishes at intervals
around $w=k+1/2$ with integer $k$.
}
\label{fig:Lin9}
\end{figure}

According to \req{Lin9c}, at sufficiently strong magnetic field
the absorption coefficient ${\cal F}_1(w,\alpha)$ vanishes at frequency
intervals such that $|w-k| > 2 \sqrt{\alpha}/\pi$ with integer $k$,
as one may expect for the case, when the density of states has gaps
between Landau levels, see e.g. \cite{Ando}.
Numerical investigation of ${\cal F}_1(w,\alpha)$ at intermediate values
of $\alpha\sim 1$ allows us to find the threshold value of magnetic field,
when the gap appears in the two level correlation function within
the SCBA; this value of the magnetic field
corresponds to $\alpha\approx 0.65$. Figure~\ref{fig:Lin9}
shows ${\cal F}_1(w,\alpha)$ for three values of $\alpha$,
including the threshold value $\alpha\approx 0.65$.
We note that the vanishing of the two level correlation function
at some energy interval is an artifact of the SCBA. However, the correction
to this result (tails in the density of states)
are known\cite{Efetov} to drop exponentially  with the
increase of the Landau level index $N$ and we disregard such tales in our
study.

\subsection{$dc$ transport at low temperature}
\label{sec:Shubnikov}
At low temperature terms in the second and third lines of \req{solf}
become important. Substitution of these terms into \req{j1} yields
for the resistivity at $\w_c\tau_{tr} \gg 1$,
compare with \req{Lin8},
\be
\begin{split}
&
\frac{\rho_{xx}(B,T)}{\rho_D}
\\
&=
\sum_{l=-\infty}^{+\infty}(-1)^l
\Upsilon\left(\frac{l T}{\hbar\w_c}\right)
\cos\left(\frac{\pi l p_F^2\lH^2}{\hbar^2}\right)
\eta_l\left(\frac{2\pi}{\w_c\tau_q}\right),
\end{split}
\mylabel{sigma2z}
\ee
where $\rho_{D}$ is the Drude resistivity, \req{rhoD}, and
\[
\Upsilon\left(x\right)=\frac{2\pi^2 x}{\sinh (2\pi^2 x)},
\quad \Upsilon\left(0\right)=1.
\]
The disorder dependent coefficients $\eta_l(\alpha)=\eta_{-l}(\alpha)$
are given by
\begin{equation}
{\eta}_l\left(\alpha\right)
=
\sum_{k=-\infty}^{\infty}
\exp\left(-\frac{\alpha(|l+k|+|k|)}{2}\right)
g_{|k|}(\alpha)g_{|l+k|}(\alpha),
\mylabel{Ak}
\end{equation}
with $g_l(\alpha)$ defined in terms of the Laguerre polynomials
by  \req{sol1}.

\begin{figure}
\epsfxsize=0.45\textwidth
\epsfbox{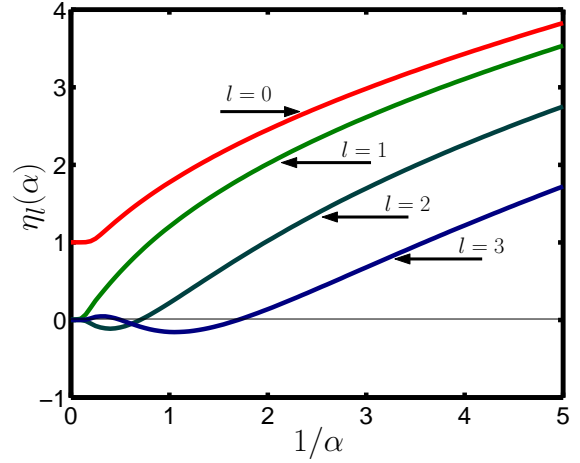}
\caption{
The harmonics $\eta_l$ of Shubnikov-de Haas  oscillations, \req{sigma2z}
as functions of magnetic field and the quantum scattering time,
$\alpha=\frac{2\pi}{\w_c\tau_q}$.
}
\label{fig:Ak}
\end{figure}

Term with $l=0$ in \req{sigma2z} reproduces the smooth part
of the magnetoresistance \rref{Lin8} and $l \geq 1$ represent
the Shubnikov-de Haas oscillations. The asymptotic behavior
of function $\eta_l(\alpha)$ is the following. At low fields,
$\alpha \gg 1$
only the first few terms are relevant
\be
\begin{split}
&\eta_1=2e^{-\alpha/2}+2(1-\alpha)e^{-3\alpha/2}+
{\cal O}\left(e^{-5\alpha/2}\right); \\
&\eta_2= \left(3-2\alpha\right)e^{-\alpha}
+{\cal O}\left(e^{-2\alpha}\right);\\
&\eta_3=\left(4-8\alpha+3\alpha^2\right)e^{-3\alpha/2}+
{\cal O}\left(e^{-5\alpha/2}\right)\\
&\eta_l={\cal O}\left(\alpha^{l-1}e^{-l\alpha/2}\right).
\end{split}
\mylabel{Aklow}
\ee
At high magnetic field $\alpha \ll 1$, we use \req{gapprox} and
obtain for $l\sqrt{\alpha}\ll 1$
\be
\eta_l(\alpha) =
\frac{16}{3 \pi\sqrt{\alpha}}
\left[
1
 - \frac{2 l^2\alpha}{5} +
{\cal O}\left(l^4\alpha^2\right)
\right].
\mylabel{Akhigh}
\ee
Coefficients $\eta_l(\alpha)$ obtained
from \reqs{Ak} and \req{sol1} are  plotted in Fig.~\ref{fig:Ak}.

Below in this paper we assume that temperature is sufficiently high,
$T\gg \w_c$. This assumption allows us
to disregard the Shubnikov--de Haas oscillations in transport
quantities.

In this Section, we applied the quantum Boltzmann
equation to calculate the linear response of electron
system on the applied electric field. The approach developed here
enabled us to describe the resistivity at arbitrary
values of the parameter $\w_c\tau_q$. Our findings are in accord with
the results of Ref.~\onlinecite{Andoreview,Ando,Ando1}
and differ by an overall numerical
factor from the corresponding result of Ref.~\onlinecite{EAltshuler}.
Particularly, the strong
magnetic field asymptote for the smooth part of the
magnetoresistance matches the result of Ref.~\onlinecite{Ando}. The
amplitude of the Shubnikov--de Haas oscillations calculated in this
Section is consistent with the previous analysis of magnetoresistance
oscillations in Refs.~\onlinecite{Ando1,EAltshuler}. We also
derived an expression for the absorption rate of microwave
radiation, \req{Lin9}. Asymptotes of our expression in weak
and strong magnetic fields coincide with the results
presented in Ref.~\onlinecite{Mirlin-ac}.
Having made sure that the consequences of the quantum Boltzmann
equation are consistent with the results obtained by different methods,
we will proceed with  the description of electron transport
beyond the linear response.

\section{Non-linear $dc$ effects}
\label{nlinear}

A strong $dc$ electric field produces non-linear effects on
(i) the even harmonics of the distribution function, see \req{mf7};
and (ii) the elastic scattering processes (spectrum),
see non-linear terms in \req{mf8}. The first mechanism roughly
corresponds to the heating and
it strongly affects system properties determined directly by
the electron distribution function, such as Shubnikov-de Haas
oscillations of the conductivity, see Sec. \ref{sec:Shubnikov}.
As we have already noticed in Sec.~\ref{sec:df}, the distribution function
is very sensitive to the details of inelastic processes.
If, however, the temperature is large,
\begin{equation}
T\gg \hbar\w_c,
\mylabel{highTlimit}
\end{equation}
from the very beginning we
can restrict our consideration to the non-linear effects
on the electron scattering process. In the remainder of the paper
we consider only the high temperature limit \req{highTlimit}.

Similarly to \req{solf}, we take into account
only collision term \rref{mf8}.
Neglecting exponentially small terms and assuming $\w_c\tau_{tr}\gg 1$,
we solve \req{mf1} and obtain
\be
\begin{split}
&2\pi \delta f\left({\cal T}^lt,t; \varphi\right)
=
\frac{p_F\partial_\varphi^{-1}}
{\w_c\tau_{tr}}
\Bigg[\lambda_l^* g^R_l(\varphi)
C\left({\cal T}^lt,t;\varphi\right)\Bigg]
\\
&C\left(t_1,t_2;\varphi\right)
=\frac{\partial}{\partial t_2}
\left[\bzeta_{12}\cdot\nn(\varphi)
h_2\left(\frac{\bzeta_{12}\hat\ep \nn(\varphi)
}{\xi}\right)
\right].
\end{split}
\mylabel{nlin1}
\ee
We notice, that the distribution function
$\delta f({\cal T}^lt,t; \varphi)$
does not depend on time $t$.
In the linear response regime $h_2=1$, see
\req{h2}, and \req{nlin1} reduces
to the first term in \req{solf}.

Using \req{j1} and  $N_e=p_F^2/2\pi$ we find
\be
\begin{split}
&\j^{(d)}=\frac{2 e N_e}{\w_c\tau_{tr}}
\int\limits_0^{2\pi} \frac{d\varphi}{2\pi}\nn(\varphi)
\Bigg\{\partial_\varphi^{-1}
C\left(t,t;\varphi\right)
\\
&
+  2 \sum_{l=1}^{\infty}|\lambda|^{2l}
g^R_{l}\left(\hat{\cal T}^lt; \varphi\right)
\partial_\varphi^{-1}
\left[
g^R_{l}\left(\hat{\cal T}^lt; \varphi\right)
C\left({\cal T}^lt,t;\varphi\right)
\right]
\Bigg\}.
\end{split}
\mylabel{nlinj1}
\ee
Equation \rref{nlinj1} is more complicated than its linear
response counterpart, because the spectrum, determined by $g^R_l(t,\varphi)$,
depends on the applied $dc$ field.
It prevents one from using \req{sol1}; therefore \req{trgr2} should be solved
again.
Using \req{smallsigma4},
and  \req{zeta} for the $\pt \E=0$,
we obtain instead of \req{trgr3}
\be
\begin{split}
&{\cal G}^R_l\left(t,\varphi\right)
=g^R_{l}(\varphi)
\\
&
 -\frac{1}{\tau_q}\sum_{m=1}^l\int_{0}^{t}dt_1 g^R_{m}
h_1\left(\frac{m\E\cdot \nn(\varphi+t_1\w_c)}{E_0}
\right)
{\cal G}^R_{l-m}\left(t_1,\varphi \right),
\\
&
g^R_{l} =
{\cal G}^R_l\left(0\right)={\cal G}^R_{l-1}\left(\frac{2\pi}{\w_c}\right);
\end{split}
\mylabel{trgr4}
\ee
where $h_1$ is defined in \req{h1}, and we
introduced a scale for electric field
\be
E_0=\frac{m_e\w_c^2\xi}{2\pi e}.
\mylabel{E0}
\ee
Explicit angular dependence in \req{trgr4} makes the
solution for an arbitrary magnetic field difficult. We
consider only limiting cases of weak, $\w_c\tau_q\ll 1$,
and strong, $\w_c\tau_q \gg 1$, magnetic fields.

At weak magnetic field, $\w_c\tau_q
\ll 1$ we can keep only  first two non-trivial terms
in \req{nlinj1}. Solutions of \req{trgr4} for $l=1,2$
are angular independent:
\[
g_1=g_0=1;\quad g_2=1-\frac{2\pi}{\w_c\tau_q}
\left(\frac{E_0^2}{E^2+E_0^2}\right)^{1/2}.
\]
Substituting these functions into \req{nlinj1} we obtain
the solution in the form
\be
\j^{(d)}
 = \frac{\E}{|\E|}
\left(\frac{e^2N_e E_0}{m_e\w_c^2\tau_{tr}}\right)
{\cal F}_2 \left( \frac{E}{E_0},\frac{2\pi}{\w_c\tau_q} \right),
\label{nlinj1aA}
\ee
where the dimensionless function ${\cal F}_2(x,\alpha)$
in the weak magnetic field can be expanded as
\be
\begin{split}
& {\cal F}_2 \left( x,\alpha \gg 1 \right)=
x \Bigg\{
1 +
\frac{2\left(1-{2x^2}\right) e^{-\alpha}}
{\left(1+{x^2}\right)^{5/2}}
\\
&
+
\frac{2\left({1}-{8x^2}\right) e^{-2\alpha}}
{\left(1+{4x^2}\right)^{5/2}}
\left(1-
\frac{\alpha}{\left(1+x^2\right)^{1/2}}
\right)^2
+{\cal O}\left[e^{-3\alpha}\right]
\Bigg\}.
\end{split}
\mylabel{nlin2}
\ee
For the weak $dc$ fields, $|\E| \ll E_0$,
\req{nlin2} match \req{Lin8a}.

At strong magnetic field $\w_c\tau_q\gg 1$ the second angular
harmonics in the solution of \req{trgr4} is suppressed by a factor
of $\alpha=2\pi/(\w_c\tau_q)$ in comparison with the zero angular harmonics.
Neglecting this correction and introducing
\begin{equation}
\tilde g_l^R(E/E_0,\alpha)=g_l^R(E/E_0,\alpha) e^{-\alpha l/2}
\mylabel{tildeg}
\end{equation}
we obtain from \req{trgr4}
\be
\tilde g^R_{l+1}(x,\alpha)
=\left[1-\frac{\alpha}{2}\right]\tilde g^R_{l}(x,\alpha)
 -\alpha \sum_{m=1}^l
\frac{\tilde  g^R_{m}(x,\alpha) \tilde g^R_{l-m}(x,\alpha) }
{\left(1+m^2x^2\right)^{1/2}}.
\mylabel{trgr5}
\ee
Equation \rref{nlinj1} simplifies to
\be
{\cal F}_2(x,\alpha \ll 1)  =  x\left[1+
2\sum_{l=1}^\infty
\frac{\left(1-2l^2x^2\right)}
{\left(1+l^2{x^2}\right)^{5/2}}\tilde g_{l}^2(x,\alpha)
\right].
\mylabel{nlinj1aB}
\ee

If the electric field is weak,
$|\E|\ll \frac{E_0}{\left(\w_c\tau_q\right)^{1/2}}$, one can
find a solution of \req{trgr5} as a correction to \req{sol1}
and use approximation similar to \req{gapprox}:
\be
\begin{split}
&\tilde g_l(x,\alpha)=
\sqrt{\frac{1}{\alpha l^2}}
J_1\left(2 \sqrt{\alpha l^2}\right)
\Gamma_1\left( lx; \alpha l^2\right);\\
&\Gamma_1(x,y)=
\left\{
\begin{matrix}
1,& y \ll 1 ;\\
\frac{\sqrt{1+x^2}}{2\sqrt{1+x^2}-1},& y \gg 1.
\end{matrix}
\right.
\end{split}
\mylabel{sol1E}
\ee
Substituting this expression into \req{nlinj1aB}, we find with the
logarithmic accuracy for $x\ll 1$
\be
\begin{split}
{\cal F}_2(x,\alpha) &=\frac{x}{\pi \sqrt{\alpha}}
\left[
\frac{16}{3}
- \frac{11x^2}{4\alpha}
\ln \left( \frac{\alpha}{x^2}\right)
\right].
\end{split}
\mylabel{nlin3}
\ee
The second term in \req{nlin3} represents the suppression of the
renormalized transport time due to the applied electric field, compare to
\req{Lin8b}.

At strong electric field, $|\E|\gg \frac{E_0}{\left(\w_c\tau_q\right)^{1/2}}$,
\req{trgr5} can be solved by perturbation theory in $1/(\w_c\tau_q)$
\be
\tilde g^R_{l}(x,\alpha)
= 1- \alpha
\left(\frac{l}{2}+
\sum_{m=1}^{l-1}\frac{(l-m) }{\left(1+m^2x^2\right)^{1/2}}
\right).
\mylabel{trgr5a}
\ee
This gives with the help of \req{nlinj1aB} and Poisson
summation formula
\be
{\cal F}_2(x,\alpha)
=
\frac{2\pi\alpha}{x^2}
+
\frac{8\pi^2}{x^{3/2}}e^{-\frac{2\pi}{x}}, \quad \sqrt{\alpha}\ll x \ll
1.
\mylabel{nl10}
\ee
For the strongest fields $|E| \gg E_0$ the main contribution to
the current becomes linear in field:
\be
{\cal F}_2(x,\alpha)
=x -4\frac{\zeta(3)-2\alpha\zeta(2)}{x^2},\quad x\gg 1,
\mylabel{nl11}
\ee
where $\zeta(x)$ is the $\zeta$-function.

\begin{figure}
\epsfxsize=0.45\textwidth
\epsfbox{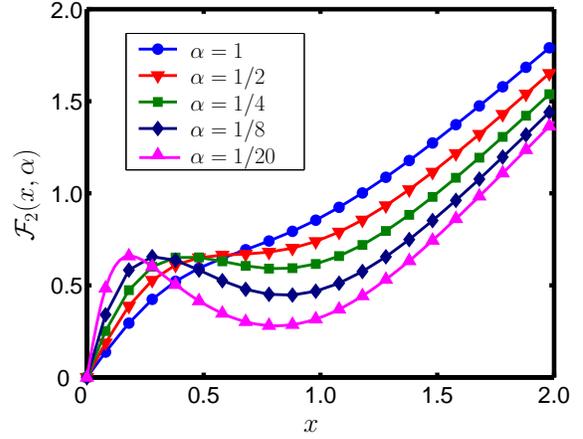}
\caption{
Nonlinear dependence of the
dissipative current on the applied
electric field $x=E/E_0$
at high-temperature $\hbar\w_c\ll T$,
for different values of the magnetic fields,
$\alpha=2\pi/\w_c\tau_q$.
}
\label{fig:nlin1}
\end{figure}

Dependence of  ${\cal F}_2(x,\alpha)$ on $x$ is plotted in
Fig.~\ref{fig:nlin1} for several values of $\alpha$.
Function ${\cal F}_2(x,\alpha)$ is calculated according
to \req{nlinj1aB} with functions $\tilde g_l(x,\alpha)$
obtained from recursive equation \req{trgr5}. At strong magnetic
field, ${\cal F}_2(x,\alpha)$ exhibits a non-monotonic behavior with
a minimum at $E\sim E_0$. At strong electric field, $E\gg E_0$,
all curves approach the zero magnetic field result,
${\cal F}_2(x,\alpha)=x$, since the strong electric field destroys
the interference effect of electron motion along cyclotron orbits.

\begin{figure}
\epsfxsize=0.45\textwidth
\epsfbox{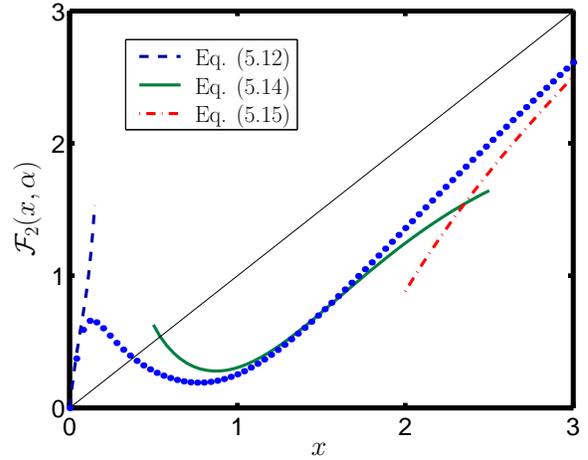}
\caption{
The plot shows three asymptotes of function ${\cal F}_2(x,\alpha)$ for
$\alpha=1/40$:
i) for small $x\ll 1$ [\req{nlin3}, dashed line];
ii) for $\sqrt{\alpha}\ll x \ll 1$ [\req{nl10}, solid line];
iii) for large $x\gg 1$ [\req{nl11}, dot-dashed line].
The dotted line represents function ${\cal F}_2(x,\alpha)$ at $\alpha=1/40$
calculated directly from \req{nlinj1aB}.
}
\label{fig:nlin1a}
\end{figure}

Figure \ref{fig:nlin1a} shows the asymptotes of function
${\cal F}_2(x,\alpha)$ for three intervals of the strength of
electric field, see \reqs{nlin3}, \rref{nl10} and \rref{nl11}. For
comparison, we also show ${\cal F}_2(x,\alpha)$ calculated
directly from \req{nlinj1aB} for $\alpha=1/40$.

\section{Effect of microwave radiation on $dc$ transport}
\label{mwdc}

Consider the two-dimensional electron gas in a magnetic
field $\w_c\tau_{tr} \gg 1$, subjected to a
monochromatic microwave radiation
together with the $dc$ field
\be
\E(t)=\E + \Re \E_{\w} e^{-i\w t},
\mylabel{mw1}
\ee
where $\E_\w$ is a complex vector in the plane of two-dimensional
electron gas.

For the strong magnetic field such as  the filling
factor is small, $\nu \simeq 1$, the effect of microwave
radiation was considered in detail
in Ref.~\onlinecite{Ryzhii}, and the linear response for the short range
disordered was analyzed in Ref.~\onlinecite{Durst}.
Our goal is to extend
these studies to the small angle impurity scattering and
to the non-linear $dc$ response. The first direction will make
the theory more adequate for the description of the experiments
\cite{Zudov1,Zudov2,Mani,Zudov4}, whereas the second
development provides the microscopic grounds of the
theory of zero-resistance state \cite{Andreev}. The latter
issue is analyzed in further details in the subsequent section.

To characterize the microwave power in
dimensionless units, we introduce
\be
{\cal P}=
\frac{\w_c^2\left(\w_c^2+\w^2-2\w\w_c\cos 2\beta\right)}
{2\left(\w^2-\w_c^2\right)^2}
\left(\frac{\w_c}{\pi\w }\right)^2
\frac{\E_\w\cdot\E_\w^*}{E_0^2},
\mylabel{mw2}
\ee
where characteristic field, $E_0$, is defined in \req{E0}.

The polarization of the microwave is described by the
angle $\beta$ and the unit vector $\mm$ as prescribed by \req{wf9}.
We will introduce also the parameter
\be
\gamma(\w)=\arctan\left(\frac{\w+\w_c}{\w-\w_c}\tan\beta\right).
\mylabel{wf10}
\ee
which describes an elliptic trajectory of a classical electron
in the magnetic and microwave fields.

The only difference of \req{mw3} from \req{nlin1} is the
time dependence of the distribution
function and the spectrum due to the oscillating microwave field.
Working in the approximation of
large Hall angle $\w_c\tau_{tr} \gg 1$ and large temperature
$T\gg \hbar\w_c$, we once again solve \req{mf1} with the collision
term \rref{mf8}. We obtain
\be
\begin{split}
& \delta f\left({\cal T}^lt,t; \varphi\right)
=
\frac{p_F(2\pi\tau_{tr})^{-1}}
{\left(\pt+\w_c\partial_\varphi\right)}
\left[\lambda_l^* g^R_l(t,\varphi)
C\left({\cal T}^lt,t;\varphi\right)\right]
\\
&C\left(t_1,t_2;\varphi\right)
=\frac{\partial}{\partial t_2}
\left[\nn(\varphi)\bzeta(t_1,t_2)
h_2\left(\frac{\bzeta(t_1,t_2)\hat\ep \nn(\varphi)
}{\xi}\right)
\right],
\end{split}
\mylabel{mw3}
\ee
where $\bzeta(t_1,t_2)\equiv \bzeta(t_1)-\bzeta(t_2)$ with
vector $\bzeta$ defined in \req{zeta}, and $\lambda$
is introduced in \req{lambda}.
Explicitly,
\[
\begin{split}
\frac{\bzeta(t_1,t_2)}{\xi}&= \frac{\w_c(t_2-t_1)
\hat\ep \E}{2\pi
E_0}+ \sqrt{{\cal P}} \sin\frac{\w(t_2-t_1)}{2}
\\
&\times\Re\left[\hat\ep
\left(\mm_+\cos\gamma+\mm_-\sin\gamma\right)e^{\frac{-i\w(t_1+t_2)}{2}}
\right]
\end{split}
\]

The spectrum of the system depends both  on the microwave radiation
and the applied electric field. From \reqs{trgr2}, \rref{gsmall2},
and \rref{smallsigma4} we find similarly to \req{trgr4}
\begin{widetext}
\be
\begin{split}
g^R_{l}(t) &=
{\cal G}^R_l\left(t,t\right)={\cal G}^R_{l-1}
\left(\hat {\cal T}^{-1}t,t\right),
\\
{\cal G}^R_l\left(t,
t_1,\varphi\right)
&=g^R_{l}(t,\varphi)
 -\frac{1}{\tau_q}\sum_{m=1}^l\int_{0}^{t-t_1}dt_2
 g^R_{m}(t_2,\varphi+t_2\w_c)
{\cal G}^R_{l-m}\left(t_2, t, \varphi \right)
h_1\left(Z_{l,m}(t_2,t_1)\right),
\\
Z_{l,m}(t_2,t_1)&=
\nn(\varphi+t_2\w_c)\left[\frac{m\E}{E_0}+
\sqrt{{\cal P}}
\sin\frac{\pi m \w}{\w_c}
\Re \left(\mm_+\cos\gamma+\mm_-\sin\gamma\right)
e^{-i\w (t_1+t_2)}e^{\frac{i\pi (m-2l)\w}{\w_c}}
\right],
\end{split}
\mylabel{trgr6}
\ee
where $h_1$ is defined in \req{h1},
time shift operator is given by \req{tshift},  field $E_0$ is defined in
\req{E0}, dimensionless power
of the microwave radiation ${\cal P}$ is given by
\req{mw2}, and the angle $\gamma$ is given by \req{wf10}.

We substitute the electron distribution function \req{mw3}
into  \req{j1} and obtain the following
expression for the $dc$ component of the electric current
in terms of $g_l^R(t,\varphi)$,
compare to \req{nlinj1}:
\be
\begin{split}
&\j^{(d)}=\frac{2 e N_e}{\w_c\tau_{tr}}
\int\limits_0^{2\pi} \frac{d\varphi}{2\pi}\nn(\varphi)
\Bigg\{\partial_\varphi^{-1}
\langle C\left(t,t;\varphi\right)\rangle_t
+  2 \sum_{l=1}^{\infty}|\lambda|^{2l}
\Bigg\langle
g^R_{l}\left(\hat{\cal T}^lt; \varphi\right)
\left(\pt+\w_c\partial_\varphi\right)^{-1}
\left[
g^R_{l}\left(\hat{\cal T}^lt; \varphi\right)
C\left({\cal T}^lt,t;\varphi\right)
\right]
\Bigg\rangle_t
\Bigg\}
,
\end{split}
\mylabel{mwj1}
\ee
\end{widetext}
and $\langle\dots\rangle_t$ stand for the time averaging over
the period of the microwave field.

Equation \rref{mwj1} together with the recursion relation
\req{trgr6} determines the electric current to all orders both in
the microwave power ${\cal P}$ and $dc$ electric field
${\bf E}/E_0$. We remind that our consideration
is valid
for large filling factors $\nu \gg 1$, large Hall angle
$\w_c\tau_{tr} \gg 1$,
large temperatures $T\gg \hbar\w_c$,
and under the conditions of the applicability of  self-consistent
Born approximation \rref{SCBAvalid}.
Further simplifications are possible for certain limiting cases, which
will be considered below.

\subsection{Weak magnetic field, $\w_c\tau_q \ll 1$}
In this case we can limit ourselves with
only first non-trivial term in \req{mwj1}. Because $g_1=1$, further
calculation is reduced to straightforward angular and time integration
in \req{mwj1}. The regimes where the compact analytic
results are available are listed below.

\subsubsection{Circular polarized microwave radiation}
\label{spmwr}

For the linear response in $dc$ electric field ${\bf E}$,
we find
\be
\j^{(d)}_c
=
\left(\frac{e^2N_e\E}{m_e\w_c^2\tau_{tr}}\right)
{\cal F}_3\left({\cal P},\frac{\w}{\w_c},\frac{2\pi}{\w_c\tau_q}\right),
\mylabel{F3}
\ee
where dimensionless microwave power is defined in \req{mw2} and
\be
\begin{split}
{\cal F}_3&({\cal P},w,\alpha)=
1 +
\frac{e^{-\alpha}
\left(2- {\cal P} \sin^2\pi w\right)}
{\left(1+{\cal P}
\sin^2 \pi w\right)^{5/2}
}
\\
&
-\frac{3 \pi w}{2}{\cal P}e^{-\alpha}\sin 2\pi w
\frac{
\left(4-   {\cal P} \sin^2\pi w\right)}
{\left(1+{\cal P}
\sin^2 \pi w \right)^{7/2}
} +{\cal O}(e^{-2\alpha }).
\end{split}
\mylabel{wmf6}
\ee
At ${\cal P}=0$, \req{wmf6} matches \req{Lin8a}.

\begin{figure}
\epsfxsize=0.45\textwidth
\epsfbox{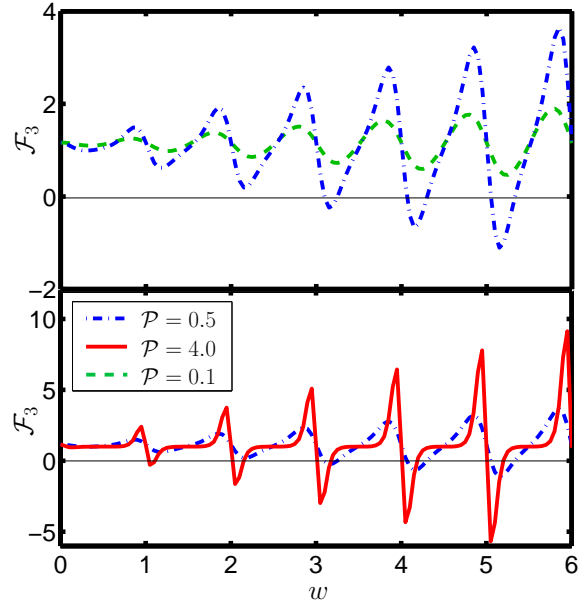}
\caption{
Frequency ($w=\w/\w_c$) dependence
of ${\cal F}_3({\cal P},w,\alpha)$
at fixed value of the microwave
power parameter ${\cal P}$ from \req{mw2}
[notice, that it corresponds
to the actual microwave power dependent on the frequency $\w$]
in the weak magnetic field $\w_c\tau_q=0.8\pi$.
The upper panel presents curves corresponding to the
weak microwave field, ${\cal P}\lesssim 1$, whereas the solid
curve in the lower panel corresponds to the strong microwave radiation
${\cal P}=4$.
Curves are calculated for the circularly polarized microwave.
}
\label{fig:mw1}
\end{figure}

Structure of \req{wmf6} deserves some additional discussion.
The second term in brackets describes the effect of the microwave on
the elastic scattering process. In the region of the applicability
of the theory $\w_c\tau_q \ll 1$, its value can never become larger
than the first term and their sum is  always positive.
This  follows from the fact
that the elastic transport cross-section is
positive by construction no matter what kind of
renormalization it acquires. The third term is the photovoltaic effect
discussed in Sec.~\ref{qd}. Its sign depends on the frequency of
the radiation $\w$ and, remarkably, on the power of the microwave
radiation.
It is noteworthy, that this term may make the current flow opposite to
the electric field even  at small magnetic field due to the presence
of possibly large factor $6\pi\w/\w_c$ in front.
Finally, we emphasize non-monotonic dependence of the
photovoltaic effect on the microwave power.
\footnote{
Regime of parameters where  \req{wmf6} is valid was studied for the short range
disorder in Ref.~\onlinecite{Durst}. Because, the final analytic
results from this reference are not available we were not able
to compare them with ours.}
The frequency dependence of the $dc$ resistivity at weak field is
plotted in Fig.~\ref{fig:mw1}.
The corrections to the Hall coefficient are small as
$1/(\w_c\tau_{tr})^2$
and will net considered here explicitly.

If the microwave power is small
\[
{\cal P} \sin^2\frac{\pi\w}{\w_c}\ll 1,
\]
one can expand \req{wmf6} up to the first order in ${\cal P}$.
In this case, the whole non-linear $dc$ response affected by
the microwave can be found, compare with \req{nlinj1aA}
\be
\j^{(d)}_c
=
\frac{\bf E}{|\bf E|}\frac{e^2N_e E_0}{m_e\w_c^2\tau_{tr}}
{\cal F}_4
\left(\frac{|\bf E|}{E_0},{\cal P},
\frac{\omega}{\w_c},\frac{2\pi}{\w_c\tau_q}\right),
\mylabel{mwnlin1}
\ee
where
\be
\begin{split}
{\cal F}_4(x,{\cal P},w,\alpha\gg 1)&
=x
\Bigg[ 1
+
\frac{2
\left(1-{2x^2}\right) e^{-\alpha}}
{\left(1+{x^2}\right)^{5/2}}
\\
&
- 3 {\cal P} e^{-\alpha}
\sin^2 \pi w
\frac{
4 -27 x^2+7 x^4}
{2\left(1+ {x^2}\right)^{9/2}}
\\
&
-3 \pi w {\cal P}e^{-\alpha}
\sin 2\pi w
\frac{  4-x^2}{2 \left(1+x^2\right)^{7/2}}
\\
&
+{\cal O}({\cal P}^2e^{-\alpha})+
{\cal O}(e^{-2\alpha})
\Bigg]
.
\end{split}
\mylabel{F4a}
\ee

\begin{figure}
\epsfxsize=0.45\textwidth
\epsfbox{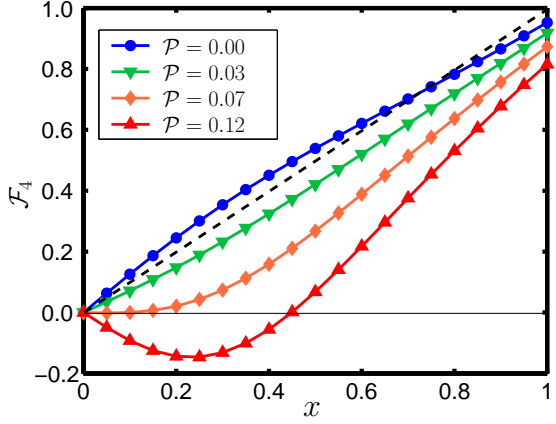}
\caption{
Non-linear dependence of ${\cal F}_4(x,{\cal P},w,\alpha)$
on the strength $x=|{\bf E}|/E_0$ of the $dc$ field
for different values of power parameter ${\cal P}$
and for $\w=7.25\w_c$. Compare with Figs.~\ref{fig:nlin1} and \ref{fig:mw1}.
Curves are plotted for
the regime of weak magnetic field $\w_c\tau_q=\pi$
and for circularly polarized microwave.
}
\label{fig:mw2}
\end{figure}

Function ${\cal F}_4(x,{\cal P},w,\alpha)$  is plotted in Fig.~\ref{fig:mw2}.
One can see, that at large electric field $x\gg 1$ the Ohm law is restored
and the microwave radiation becomes irrelevant in accord
with the conjecture of Ref.~\onlinecite{Andreev}.
We will discuss consequences of negative values of ${\cal F}_4$
in Sec.~\ref{domain}.

At weak $dc$ field, $|E| \ll E_0$
one can keep only terms linear in $dc$ field
in \req{mwnlin1},
\be
\begin{split}
\j^{(d)}_c
&=
\left(\frac{e^2N_e\E}{m_e\w_c^2\tau_{tr}}\right)
\Bigg\{
1 +2 e^{-\frac{2\pi}{\w_c\tau_q}}
\\
&
- 6{\cal P} e^{-\frac{2\pi}{\w_c\tau_q}}
\left( \sin^2\frac{\pi \w}{\w_c}
+\frac{ \pi \w\sin \frac{2\pi \w}{\w_c}}{\w_c}
\right)
\Bigg\}
.
\end{split}
\mylabel{wmf6a}
\ee
It coincides with the corresponding expansion in \req{wmf6}.

\subsubsection{Arbitrary polarization of the microwave radiation}

The compact results can be obtained for the first order expansion
in ${\cal P}\sin^2\frac{\pi\w}{\w_c} \ll 1$. Polarization
of the microwave is characterized by the parameter $\gamma$ from
\req{wf10}.
We find
\be
\j^{(d)}(\gamma)=\j^{(d)}_c + \delta\j_a \sin 2\gamma,
\mylabel{mwnlin2}
\ee
where $\j^{(d)}_c$ represents the isotropic component of the
current and coincides with the current produced by circularly
polarized microwave field, \req{mwnlin1}.
The anisotropic component is given by
\be
\begin{split}
\delta\j_a &=\frac{e^2N_e E_0}{m_e\w_c^2\tau_{tr}}
\Bigg\{
\frac{\E -2 \mm (\mm\E)}{|\E|}
{\cal F}_5^{d}
\left(\frac{|\E|}{E_0},{\cal P},\frac{\w}{\w_c},
\frac{2\pi}{\w_c\tau_q}\right)
\\
&
- \frac{\E}{|\E|} \frac{ \E^2- 2\left(\mm\E\right)^2 }{|\E|^2}
{\cal F}_5^{q}
\left(\frac{|\E|}{E_0},{\cal P},\frac{\w}{\w_c},
\frac{2\pi}{\w_c\tau_q}\right)
\Bigg\},
\end{split}
\mylabel{ja}
\ee
where functions for the dipole and quadruple angular harmonics are given by
\begin{subequations}
\label{mwnlin30}
\begin{eqnarray}
&&{\cal F}_5^d(x,{\cal P},w,\alpha)=
\frac{3 \pi w \sin 2\pi w}{ \left(1+x^2\right)^{5/2}}
+ \frac{3 (1-4x^2)\sin^2\pi\w}{ \left(1+x^2\right)^{7/2}}
;\nonumber
\\
\\
&&{\cal F}_5^q(x,{\cal P},w,\alpha)
=\frac{15}{2}\left[
\frac{\pi w\sin 2\pi w }
{ \left(1+x^2\right)^{7/2}}
+ \frac{
(3 -4x^2)\sin^2{\pi\w}
}{\left(1+x^2\right)^{9/2}}
\right] \nonumber\\
\end{eqnarray}
for ${\cal P}\ll 1,\ \alpha \gg 1$.
\end{subequations}

For the current linear in the $dc$ field and bilinear in the
microwave field,
\req{mwnlin2} simplifies to
\be
\begin{split}
\delta\j_a & =\frac{3e^2N_e{\cal P}e^{-\frac{2\pi}{\w_c\tau_q}}
}{m_e\w_c^2\tau_{tr}}\left[\E-2 \mm \left(\mm\cdot \E\right)
\right]
\\
&
\times
\left(
\frac{\pi\w}{\w_c}\sin\frac{2\pi\w}{\w_c}
+ \sin^2\frac{\pi\w}{\w_c}
\right).
\end{split}
\mylabel{mwnlin4a}
\ee
We emphasize that the anisotropy of the electric current versus
the applied electric field appears both in the
linear and non-linear $dc$ transport.

\begin{figure}
\epsfxsize=0.45\textwidth
\epsfbox{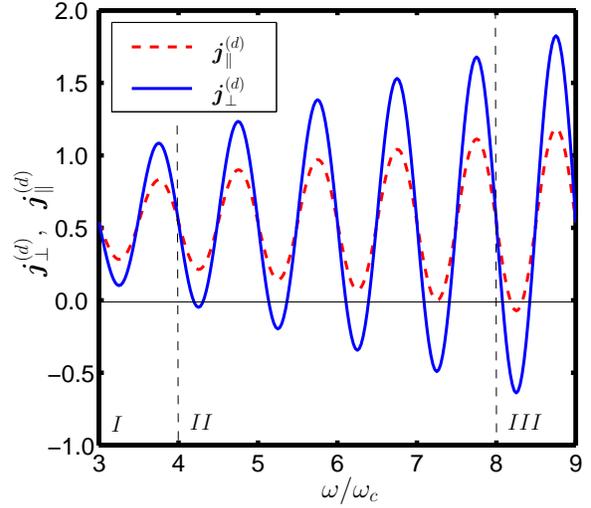}
\caption{
Dependence on microwave frequency of the electric current for a
linear polarization of the microwave field: i) $\mm\|\E$ (dashed
line) and ii) $\mm\bot\E$ (solid line). The curves are calculated
for weak magnetic field $\w_c\tau_q=\pi$ and weak power of
microwave ${\cal P}=0.2$. The strength of the electric field is
$|\E|=0.5E_0$. The plot shows that the possibility to form a
current in the direction opposite to the electric field depends on
the polarization of microwave field.
}
\label{fig:anis}
\end{figure}

In Fig.~\ref{fig:anis} we plot the $dc$ resistivity
 for the linear polarization, $\beta=\pi/4$,
of microwave field for the cases $\E\|\mm$ and  $\E\bot \mm$.
One can see from Fig.~\ref{fig:anis} that
the condition for
the electric current to flow against the applied electric field $\E$
depends on the polarization of the microwave field with respect to
the direction of $\E$.

\subsection{Strong magnetic field, $\w_c\tau_q\gg 1$}

In this case, \req{trgr6} can also be significantly simplified.
We will limit ourselves with the first order expansion
in microwave power ${\cal P}$.

First we analyze  the first order  correction in ${\cal P}$ to the
spectral functions $g_l^R$ in \req{trgr6}.
By inspection, one can see that this equation contains
terms either slowly changing during the cyclotron period or
oscillating with frequencies $n\w$.
The oscillating with frequency $2\w$ term do not contribute
at all to the final answer, whereas the term oscillating
with frequency $\w$ can be taken into account perturbatively.
Thus, for the redefined spectral functions according \req{tildeg},
we obtain analogously to \req{trgr5}
\be
\tilde g^R_{l}(t,\varphi)=\tilde g^R_{l}+\Re \delta \tilde g_l(\varphi)
e^{-i\w t-i\frac{2\pi l\w}{\w_c}},
\mylabel{trgr8}
\ee
where the angle independent component $\tilde g^R_{l}$
satisfies the following recursion relation
\be
\begin{split}
\tilde g^R_{l+1}
= & \tilde g^R_{l}\left[1-\frac{\pi}{\w_c\tau_q}\right]
\\
&
 -\frac{2\pi}{\w_c\tau_q}\sum_{m=1}^l
\frac{E_0 \tilde g^R_{l-m} \tilde g^R_{m}}
{\left( E_0^2+ m^2E^2 \right)^{1/2}}
\Bigg\{1 - {\cal P}\Xi_1(m)\Bigg\},
\end{split}
\mylabel{trgr7}
\ee
and the angle dependent component $\delta g^R_{l+1}(\varphi)$
can be found from
\be
\delta g^R_{l+1}(\varphi)
=\delta g^R_{l}(\varphi)
 +\frac{2\pi\sqrt{\cal P}e^{\frac{2\pi i\w\varphi}{\w_c}} }{\w_c\tau_q}
\sum_{m=1}^l\Xi_2(m) g^R_{l-m} g^R_{m}
.
\mylabel{trgr7a}
\ee
The initial conditions for the recursion relations are
$g^R_{0} = 1,\  \delta g^R_0=0$.
Above we introduced the  short-hand notations
\be
\begin{split}
\Xi_1(m)&=
\sin^2\frac{\pi m\w}{\w_c} E_0^2\\
&\times
\frac{
{2E_0^2}-{m^2\E^2}
-3m^2 \sin 2\gamma \left[2(\mm \E)^2-\E^2\right]
}{4 \left(E_0^2+ m^2\E^2 \right)^2}
\end{split}
\mylabel{gamma1m}
\ee
and
\be
\begin{split}
&\Xi_2(m)=
 \sin \frac{\pi m\w}{\w_c} E_0^4\int\limits_0^{2\pi}\frac{d\phi}{\pi}
\exp\left(-\frac{2\pi i\w\phi}{\w_c}\right)
\\
&\times
\frac{  \left[m \E\cdot \nn(\phi)\right]
\left[\left(\mm_+\cos\gamma+\mm_-\sin\gamma\right)\cdot \nn(\phi)\right]
}{\left(E_0^2+ \left(m\nn(\phi)\cdot\E\right)^2\right)^{2}},
\end{split}
\mylabel{gamma2m}
\ee
where $\gamma $ is given by \req{wf10}.

\begin{figure}
\epsfxsize=0.45\textwidth
\epsfbox{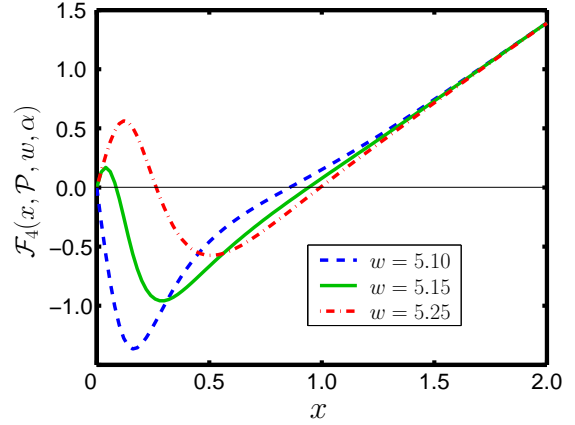}
\caption{
Nonlinear dependence of ${\cal F}_4$, \req{mwnlin1a}, on
the strength $x=|\E|/E_0$ of the $dc$ electric field at
several values of the microwave frequency $w=\w/\w_c$ in strong
magnetic field $\w_c\tau_q=40\pi$ at ${\cal P}=0.05$.
}
\label{fig:mw3}
\end{figure}

One can see, that due to the oscillating factor $
e^{\frac{2\pi i\w\varphi}{\w_c}} $ in \req{trgr7a} the contribution of
this term is suppressed by additional factor of
$\w_c/\w$ in comparison with contribution of $\Xi_1$ in
\req{trgr7}. Nevertheless, even $\Xi_1$, which describes
the effect of the microwave radiation on
the density of states, is suppressed in comparison with
photovoltaic effect by a factor of $\w_c/\w$.
Thus, in the consideration of the transport
at
\be
\w \gg \w_c,
\mylabel{largeomega}
\ee
we replace $\tilde g_l^R(t,\varphi)$, defined by \req{trgr7},
with $\tilde g_l^R(t,\varphi)$ obtained from \req{trgr5}.
All the further formulas of this Section are valid in
this high-frequency limit only.

We present the $dc$ current in the form
of \req{mwnlin2}, where the polarization dependence is
characterized by factor $\gamma$, \req{wf10}.
Keeping in mind condition \rref{largeomega}, we obtain from
\reqs{mw3} and \rref{mwj1}
 the following expressions for the functions defined in \reqs{mwnlin1}
and \rref{ja}
\be
\begin{split}
& {\cal F}_4(x,{\cal P}\ll 1,w \gg 1,\alpha \ll 1)
\\
&
\ \ = {\cal F}_2(x,\alpha)
 -
 \frac{3\pi x w{\cal P}}{2}
\sum_{l=1}^\infty
\frac{\left(4-l^2x^2\right) \sin (2\pi l w)}{(1+l^2x^2)^{7/2}}
\tilde g_{l}^2(x,\alpha),
\end{split}
\mylabel{mwnlin1a}
\ee
and
\begin{subequations}
\label{mwnlin3a}
\begin{eqnarray}
&&{\cal F}_5^d(x,{\cal P},w,\alpha)=
{3\pi w}{\cal P}\sum_{l=1}^{\infty}
\frac{l\sin 2\pi l w}{\left(1+l^2 x^2\right)^{5/2}}
\tilde g_l^2(x,\alpha);
\nonumber\\
\\
&& {\cal F}_5^q(x,{\cal P},w,\alpha)=
\frac{15\pi w}{2}{\cal P}\sum_{l=1}^{\infty}
\frac{l^3\sin 2\pi l w}{\left(1+l^2 x^2\right)^{7/2}}
\tilde g_l^2(x,\alpha).
\nonumber\\
\end{eqnarray}
\end{subequations}
for ${\cal P}\ll 1,\ w\gg 1, \alpha \ll 1$
Here, the spectral functions $\tilde g_l$ are solutions of \req{trgr5}.

In Fig.~\ref{fig:mw3} we present ${\cal F}_4(x,{\cal P},w,\alpha)$
as a function of the strength of the electric field $x=|\E|/E_0$
for several values of frequency $w=\w/\w_c$ at strong magnetic field,
$\w_c\tau_q=40\pi$. We observe that the effect of microwave field
on $dc$ current is significant only in the non-linear region,
$|\E|\lesssim E_0$. At stronger electric fields $|\E|\gg E_0$, the
effect of microwave on the $dc$ current disappears, see
Sec.~\ref{qd} for a discussion.

\begin{figure}
\epsfxsize=0.45\textwidth
\epsfbox{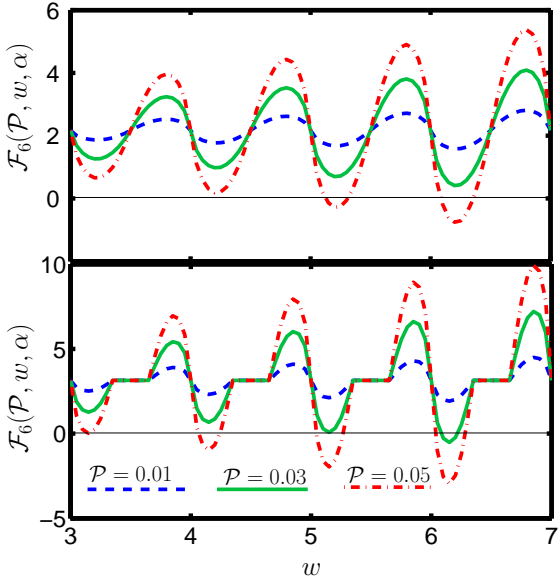}
\caption{
Dependence of
${\cal F}_6({\cal P},w,\alpha)$ on frequency $w=\w/\w_c$
for several values of ${\cal P}$. The upper panel represents the
curves calculated for $\alpha=0.65$, and the lower panel shows the
curves for $\alpha=0.30$; cf. Fig.~\ref{fig:Lin9}.
}
\label{fig:mw4}
\end{figure}

We also calculate the $dc$ current linear in the $dc$ field $\E$
and bilinear in the microwave field. For simplicity we consider
only the isotropic component of the current, which
survives at $\gamma=0$ in \req{mwnlin2} and
corresponds to the current produced by the circular polarization.
In this case we  use the spectral function $g_l(x,\alpha)$,
given by \req{sol1} and obtain
\begin{equation}
\j_c^{(d)}=\frac{e^2N_e \E}{m_e\w_c^2\tau_{tr}}
{\cal F}_6({\cal P},w,\alpha),
\mylabel{mwnlin4}
\end{equation}
where
\begin{equation}
{\cal F}_6({\cal P},w,\alpha)=
\lim_{x\to 0}
\frac{{\cal F}_4(x,\dots )}{x}=
\eta_0(\alpha)+\frac{3 w{\cal
P}}{2}\frac{\partial}{\partial w}{\cal F}_1(w,\alpha),
\mylabel{f6}
\end{equation}
and functions $\eta_0(\alpha)$ and ${\cal F}_1(w,\alpha)$ are
defined by \reqs{eta0} and \rref{Lin9a} respectively.
Relation between
the absorption spectrum and the
microwave frequency dependence of
the photovoltaic effect \rref{f6} was argued recently
in Ref.~\onlinecite{HeandShe} on the basis of a ``toy model''.
Dependence of
${\cal F}_6({\cal P},w,\alpha)$ on frequency $w=\w/\w_c$
is shown in Fig.~\ref{fig:mw4} for
several values of ${\cal P}$. [Note that fixed ${\cal P}$
corresponds to the frequency dependence of the actual microwave power,
see \req{mw2}.]

At strong magnetic field we use \reqs{Lin8b} and \rref{Lin9c} to
find the asymptotic form of function
${\cal F}_6({\cal P},w,\alpha)$ at $\alpha\ll 1$:
\begin{equation}
{\cal F}_6({\cal P},w,\alpha)=\frac{16}{3\pi\sqrt{\alpha}}
\left[
1+\frac{3\pi}{2}\frac{{\cal P}w}{\sqrt{\alpha}}
\sum_{k}
{\cal H}_2\left(\frac{\pi |w-k|}{\sqrt{\alpha}}\right)
\right],
\mylabel{f60}
\end{equation}
where for $|x|\leq 2$
$$
{\cal H}_2(x)=\frac{3x}{8}
\Bigg\{
(2+x)E\left(\frac{2-x}{2+x}\right)
-4K\left(\frac{2-x}{2+x}\right)
\Bigg\},
$$
and ${\cal H}_2(x)=0$ otherwise,
see also discussion in the last paragraph of Sec.~\ref{sec:acdc}.
Function ${\cal H}_2(x)$ has the
minimum at $x_{min}\approx 0.834$, where ${\cal H}_2(x_{min})\approx
0.726$. Correspondingly, function ${\cal F}_6({\cal P},w,\alpha)$
becomes negative if the  microwave power ${\cal P}$ exceeds
${\cal P}_{min}\approx 0.29 \sqrt{\alpha}/w$
(here $\alpha\lesssim 1$ and $w\gtrsim 1$). This expression
demonstrates that at strong magnetic fields, $\alpha\lesssim 1$,
already a weak microwave is sufficient to create
a state with zero-bias negative resistance.

\section{Formation of inhomogeneous phases and current in domains}
\label{domain}
\begin{figure}
\epsfxsize=0.45\textwidth
\epsfbox{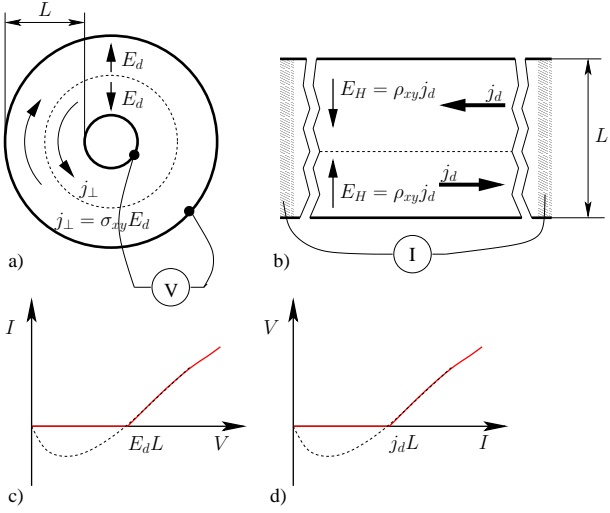}
\caption{
Domain structure\cite{Andreev} for the Corbino (a) and
Hall bar (b) geometries. For the Hall geometry, the applied current
$I < j_dL$ is accommodated by the shift of the domain
wall without any voltage drop, $V_x=0$.
At $I>j_0L$ ($V>E_d$) for the Hall bar (Corbino) geometry
domain structure is destroyed and the state with finite
dissipation is stable (c). Note, that the construction (b)
does not describe the current pattern near the leads.
}
\label{fig:ds}
\end{figure}

\begin{subequations}
\label{jvsE}

Results of the previous section qualitatively consistent
with the conclusions of Ref.~\onlinecite{Ryzhii,Ryzhii1,Durst}
indicate that there is a region in the parameter space where
the linear dissipative conductivity becomes negative.
According to Ref.~\onlinecite{Andreev}, spatially homogeneous
state of such system is unstable and break itself into the domains
characterized by zero dissipative resistivity and conductivity and
by the classical Hall resistivity,
see Fig.~\ref{fig:ds}. In the analysis of such state
one can ask two main questions: (i) what the spatial structure
of the domain wall and the boundary conditions fixing the position
and the size of the domains are; (ii) what the values of
the current and the electric field inside domains are; the
value of electric field can be found by the local probe measurement.

First question has to be answered by analyzing spatially inhomogeneous
problem by taking into account the gradient term in \req{mf8}
and the Poisson equation; this question is left for future study.
Here, we use the results of Sec.~\ref{mwdc} to address the second question.
\footnote{The non-linear effects of the electric field on
the electron distribution function, which are not taken into account,
may change the position of the boundaries between different phases of
the electron system. However, the overall structure of the phase
diagrams remains intact even if the non-linear effects are
considered.}

To clarify further consideration, let us discuss the relation
between applied current and voltage in more details.
In all of the above analysis we assumed that the electric field $\E$ is
applied and the current $\j$ is measured, the current has both the
dissipative and Hall components; the corrections to the Hall
coefficient are small as $1/(\w_c^2\tau_{tr}^2)$.
Restoring the Hall current we write the expression
for the total $dc$ current up to the terms
${\cal O}(\w_c^{-2}\tau_{tr}^{-2})$
\be
\j
 =
\left(\frac{e^2N_e E_0}{m_e\w_c^2\tau_{tr}}\right)
\hat {\cal F} \left( \frac{\E}{E_0}, {\cal P}, \frac{\w}{\w_c},
 \frac{2\pi}{\w_c\tau_q} \right)\frac{\E}{|\E|}
- \frac{1}{\rho_{xy}} \hat \ep \E,
\label{jvsEa}
\ee
where $\rho_{xy}$ is the classical Hall coefficient, see \req{Hall},
and $\hat{\cal F}$ is the tensor defined for different
situations  in \reqs{nlinj1aA}, \rref{mwnlin1}, and \rref{ja}.
[Tensor structure appears due to the microwave radiation with
polarization other than circular, see \reqs{mwnlin2} --
\rref{mwnlin4a}, \rref{mwnlin3a}.
]
Microwave power is characterized by \req{mw2} and the field $E_0$
is given by \req{E0}.
This relation is convenient to use for the Corbino disk measurement
scheme.
 For the  Hall bar geometry \req{jvsEa} can be easily inverted
\begin{equation}
\E= -{j_0}{\rho_D}
\hat\ep
\hat{\cal F} \left( \frac{\j}{j_0}, {\cal P}, \frac{\w}{\w_c},
 \frac{2\pi}{\w_c\tau_q} \right)\frac{\hat\ep\j}{|\j|}
+ {\rho_{xy}} \hat\ep \j  ,
\label{jvsEb}
\end{equation}
where $\hat\ep^2=-1$,
$\rho_D$ is the classical Drude resistivity \rref{rhoD}, and
\end{subequations}
\begin{equation}
j_0=\frac{E_0}{\rho_{xy}}=eN_e\frac{\xi\w_c}{2\pi}
\mylabel{j0current}
\end{equation}
is the electric current scale for non-linear effects.
Equations~\rref{jvsE} shows that both Hall bar and Corbino
measurements should exhibit the similar non-linear properties,
as will be  discussed below.

We mainly focus our consideration on the circular polarization of
microwave radiation; non-circular polarization is briefly discussed
in the end of this section. Then, tensor $\hat{\cal F}$ from \reqs{jvsE}
is reduced to scalar and the condition of the local stability of
the state takes the form\cite{Andreev}
\begin{subequations}
\label{stability}
\be
j_d=x j_0; \quad E_d=x E_0;
\label{stabilitya}
\ee
where $x$ is the solution of
\be
{\cal  F} \left(x, {\cal P}, \frac{\w}{\w_c},
 \frac{2\pi}{\w_c\tau_q} \right)=0;
\quad
\partial_x {\cal  F} \left(x, {\cal P}, \frac{\w}{\w_c},
 \frac{2\pi}{\w_c\tau_q} \right) > 0.
\label{stabilityb}
\ee
\end{subequations}
All the further analysis is reduced to the substituting
of appropriate limit of the function \rref{mwnlin1}
into the stability condition \rref{stability}.
\begin{figure}
\epsfxsize=0.45\textwidth
\epsfbox{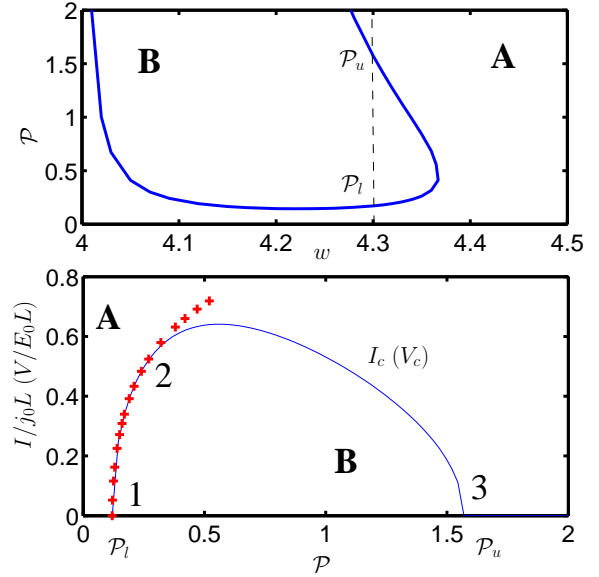}
\caption{
Upper panel:  Phase diagram of
a 2DEG in weak magnetic field ($\alpha=2$)
in ${\cal P} - w=\frac{\w}{\w_c}$
coordinates [$I=0$ ($V=0$) for the Hall bar
(Corbino) geometries].
Region (A) is the dissipative state;
region (B) is
the zero resistance (conductance) state.
Lower panel: Phase diagram for $\w=4.3\w_c$ (dashed line on the upper panel)
in ${\cal P}\! - \!{I}$
coordinates for the Hall bar (${\cal P}\! -\! {V}$ for the Corbino)
geometries. The zero resistance (conductance) state exists if the
current through the Hall bar (voltage drop between edges of the Corbino disk)
does not exceed the critical value $I_c$ ($V_c$).
The same curve defines the value of the spontaneous
current $j_d=I_c/L$ (electric field $E_d=V_c/L$) in domains.
The line labelled by ``+''  is
the  numerical solution of  \req{stability2} and the line ``2-3'' is
a schematic interpolation  beyond the linear expansion in microwave
power.
Points ${\cal P}_{u,l}$ are the same as in the upper panel.
}
\label{fig:pd}
\end{figure}

The regime of the weak magnetic field $\alpha =
\frac{2\pi}{\w_c\tau_q} \gg 1$ is  simplest. According to \req{F3},
the zero current state is stable if ${\cal F}_3>0$, thus,
the equation
\be
{\cal F}_3\left({\cal P},\frac{\w}{\w_c}, \alpha\right)=0
\mylabel{F3sta}
\ee
gives the boundary between dissipative and zero resistance state (ZRS)
for the Hall bar geometry
or zero conductance state (ZCS) for the Corbino disk geometry
in ${\cal P}\! -\! \w$ plane, where
${\cal F}_3\left({\cal P},\frac{\w}{\w_c},\alpha\gg 1\right)$
is given by \req{wmf6}. The curve given by \req{F3sta}
is plotted in the upper panel
of Fig.~\ref{fig:pd}.
For $w=\frac{\w}{\w_c}\gg 1$ the analytic estimates for the
``phase boundary'' lines are
\begin{subequations}
\begin{eqnarray}
{\cal P}_{l} & \simeq & \frac{e^{\alpha}}{6\pi w \sin 2\pi
w},\quad {\cal P}_{l}\ll 1;
\label{jdom5}
\\
{\cal P}_{u} & \simeq & \frac{4}{\sin^2\pi w}
-5^{7/2} {\cal P}_{l},\quad {\cal P}_{u}\gg 1.
\label{jdom6}
\end{eqnarray}
 \end{subequations}
The zero resistance state is impossible not only at too
low microwave power but also at excessive microwave power
(reentrance transition).
Indeed, a weak microwave radiation does not produce strong
enough photovoltaic current to compensate the dissipative current.
On the other hand, as we discussed in Sec.~\ref{qd},
strong microwave radiation suppresses the electron returns
to the same impurity and thus destroys the non-linear effects.

\begin{figure*}
\epsfxsize=0.65\textwidth
\centerline{
\epsfbox{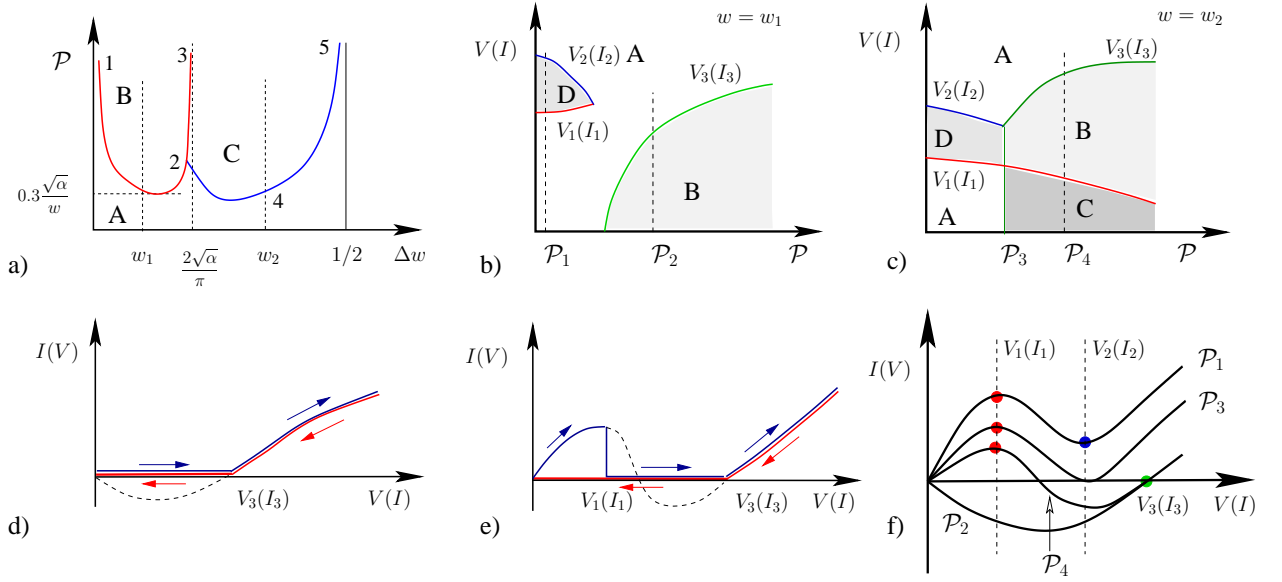}}
\caption{
(a) Phase diagram of a 2DEG in
strong magnetic field ($\alpha \ll 1$) in ${\cal P} -
w=\frac{\w}{\w_c}$ coordinates [$I=0$ ($V=0$) for the Hall bar
(Corbino) geometries]. The dissipative (A), the zero resistance
(conductance) state (B) and the coexistence regions (C) are shown.
(b,c) Phase diagrams for $\w/\w_c=w_1$ and $\w/\w_c=w_2$ (vertical dashed
lines on the upper plane) in ${\cal P}\! - \!{I}$ coordinates
for the Hall bar
(${\cal P}\! -\! {V}$ for the Corbino) geometries.
Lines $V_3(I_3)$ also describe the current in domains as functions of
the microwave power. The dissipative instability region is denoted
by D.
(d,e) The $I-V$, ($V-I$) characteristics for the
Hall bar (Corbino disk) geometries at ${\cal P}={\cal P}_2$ and
${\cal P}={\cal P}_4$ respectively; (f) Relation of the
position of the border lines to the results of Fig.~\ref{fig:mw3}.
}
\label{fig:pda}
\end{figure*}

At microwave power within the zero-resistance region
\req{stabilityb} has the solution at $x \neq 0$
\be
{\cal  F}_4 \left(x, {\cal P}, \frac{\w}{\w_c},
 \frac{2\pi}{\w_c\tau_q} \right)=0.
\label{stability2}
\ee
For the low microwave power response ${\cal F}_4$ is given by
\req{mwnlin1}.

Phase boundary given by \req{stability2} is shown on the lower
panel of Fig.~\ref{fig:pd} by the $1-2-3$ curve.
In the vicinity of the lower boundary (segment $1-2$)
and at $\w\gg \w_c$ we have
\begin{equation}
\begin{pmatrix} j_d\\ \\E_d\end{pmatrix}=
\begin{pmatrix} j_0\\ \\ E_0\end{pmatrix}
\sqrt{\frac{4\left({\cal P}-{\cal P}_{l}\right)}
{15{\cal P}_{l}}},
\end{equation}
where ${\cal P}_l$ is given by \req{jdom5}. As the power
increases, the current in domains reaches maximum and then
decreases. This non-monotonic behavior is schematically shown by
the $1-2-3$ line in the lower panel of Fig.~\ref{fig:pd}. The
corresponding segment ($2-3$) may be obtained from calculations outlined in
Sec.~\ref{spmwr} beyond the bilinear response in the microwave
field, which was not done in the present paper.
[Based on the results presented in Sec.~\ref{spmwr},
only point $3$ of this segment is known.]
However, there is no reason to expect any singular behavior of this curve.

The lower panel of Fig.~\ref{fig:pd} may be also
used as a phase diagram in ${\cal P}\ -\ I$ plane, where $I=jL$
is the total current through the Hall bar of width $L$
or in  ${\cal P}\ -\ V$ plane for the Corbino geometry, see
Fig.~\ref{fig:ds}.

We now turn to the discussion of the strong magnetic field
regime $\w_c\tau_q \gg 1$. Naively, one would expect
that increase of the magnetic field would change the phase diagram
of Fig.~\ref{fig:pd} only quantitatively by rearranging the boundary
line. However, this expectation is not correct.
We start from the phase diagram on ${\cal P} \!-\! w$ plane,
Fig.~\ref{fig:pda}a.
The condition for the boundary  between the
dissipative and ZRS (ZCS) \rref{F3sta} is modified as
\be
{\cal F}_6\left({\cal P},\frac{\w}{\w_c}, \alpha\right)=0,
\mylabel{F6sta}
\ee
where ${\cal F}_6$ is given by \reqs{f6} and \rref{f60}.
Solution of \req{F6sta} gives the line $1-2-3$ in Fig.~\ref{fig:pda}a.
On can see that the region of the instability shrinks with increasing
of the magnetic field. It is not the end of the story though.
According to Fig.~\ref{fig:mw3}, see curves for $w=5.15$ and $w=5.25$,
the state with positive zero-field
resistance but with dissipative electric field antiparallel to
the electric current at some finite
current is possible. The boundary line (curve $3-2-4-5$ in Fig.~\ref{fig:pda}a)
for such state is given
by, see also \req{stabilityb}
\bea
{\cal  F}_4 \left(x_*, {\cal P}, \frac{\w}{\w_c},
 \frac{2\pi}{\w_c\tau_q} \right)=0;\nonumber\\
\partial_x {\cal  F}_4 \left(x_*, {\cal P}, \frac{\w}{\w_c},
 \frac{2\pi}{\w_c\tau_q} \right) = 0,
\mylabel{coexistence}
\eea
where ${\cal  F}_4$ is given by \req{mwnlin1a}.
The solution of \req{coexistence} is shown as the line $2-4-5$
in Fig.~\ref{fig:pda}a. Therefore, the phase diagram becomes more
complicated. The region of the ZRS (ZCS) has the same properties
as its counterpart for the weak field. On the other hand, in the
coexistence region (C), see Fig.~\ref{fig:pda}a,c,
both the homogeneous  dissipative state with
zero current and the domain structure of the ZRS (ZCS) are locally
stable. We believe that such bistability can cause the hysteretic
behavior of the $I-V$ characteristic of the sample, see
Fig.~\ref{fig:pda}(e).

The aforementioned complication translates  into the qualitative
change in the phase diagrams in ${\cal P}\! -\! {V}$
(${\cal P}\! -\! {I}$) coordinates, see Fig.~\ref{fig:pda} (b,c)
in comparison with the lower panel of Fig.~\ref{fig:pd}.
Equations for the lines on   Fig.~\ref{fig:pda}(b,c) are
\[
\begin{matrix}
\partial_x {\cal  F}_4 \left(x, {\cal P}, \frac{\w}{\w_c},
 \frac{2\pi}{\w_c\tau_q} \right)=0; && {\rm Lines }\ \ \ V_{1,2}\ (I_{1,2})\\
{\cal  F}_4 \left(x, {\cal P}, \frac{\w}{\w_c},
 \frac{2\pi}{\w_c\tau_q} \right)=0; && {\rm Line }\ \ \ V_{3}\ (I_{3}),
\end{matrix}
\]
and $V=xLE_0$ ($I=xLj_0$).
The physical meaning of those lines is illustrated on
Fig.~\ref{fig:pda}f.

\begin{figure}
\epsfxsize=0.45\textwidth
\epsfbox{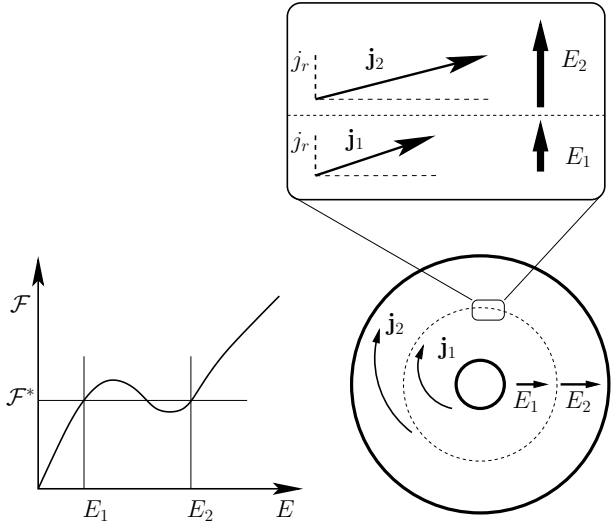}
\caption{The schematic picture of the current distribution in
the instability region D for the Corbino geometry.}
\label{fig:gunn}
\end{figure}

The region D in  Fig.~\ref{fig:pda} (b,c)
represents the state with negative differential conductance
(for Corbino geometry). In this case, the homogeneous
state is unstable, whereas the zero-resistance state is not possible.
The instability of the homogeneous state, see Fig.~\ref{fig:gunn}
leads to the formation of the domain structure with the charge
distribution similar to the Gunn domain\cite{Gunn}.
This structure will be moving from the boundary to the boundary
with the velocity $\xi\w_c{\cal F}^*/(2\pi)$, thus the domain
will be annihilated on the contact with the other one formed on
the opposite contact; so that the current pattern will
be oscillating in time rather than stationary.

Finally, let us discuss the role of the anisotropy of the dissipative
conductivity tensor in the formation of
zero-resistance (conductance) state for the linear polarization of microwave.
Here, two situations are possible, (i) both main components of the
linear resistivity tensor are negative though different;
(ii)the main components of the
linear resistivity tensor are of different signs, see
Fig.~\ref{fig:anis}.
Study of the regime (i) can be reduced to the previously studied case by
rescaling of the coordinate, currents and field, such
that equations $\nnabla\cdot \j=0$, $\nnabla\times\E=0$ are kept intact.
It does not change the state of Ref.~\onlinecite{Andreev}
qualitatively, though extra singularities may be needed to
accommodate the change in the boundary conditions. For case (ii),
the homogeneous state can be shown to be unstable, whereas the
domain structure with closed current loops would violate the condition
$\oint \E \cdot d{\bf l}=0$, because on such contour there must be
regions of positive resistance. The details of current pattern for this case
requires the further investigation, we believe, however, that the
stationary solution for this case is not possible and
domains oscillating in time will be formed.

\section{Conclusions}
In this paper,
we derived the kinetic equation within the self consistent Born
approximation for large filling factors. The obtained equation
are written in terms of the Green functions integrated in the
phase space in the direction perpendicular to the Fermi surface
similarly to the Eilenberger equation for normal metals and superconductors.
Our system of equations takes into
account the effect of electric and magnetic fields on
the  elastic scattering process, {\em i.e.} on
both the spectral function and the electron distribution function.

Armed with the quantum kinetic
 equation for the limit of
large  Hall angle, we described the following
phenomena: (i) $dc$ and $ac$ magnetoresistance in the linear response;
(ii) non-linear $dc$ current-voltage characteristic; (iii) influence of
oscillating microwave electric field on $dc$ current.
It is important to emphasize that the non-trivial
effects of the theory are described in terms of only two free
parameters, time $\tau_q$ which can be extracted for the
Shubnikov--de Haas oscillations, and the characteristic electric
field $E_0$ from \req{E0}.
The major problem of the presented paper is
the lack of the consideration of the inelastic processes and
consequently, effects related to the form of electron distribution
function. The treatment of inelastic processes will be presented
in ref.~\onlinecite{inelastic}.

We conclude by mentioning another consequence of the proposed in
Ref.~\onlinecite{Andreev}
domain mechanism of zero resistance (ZRS) and zero conductance states (ZRC).
Namely, according to
our finding, the  zero dissipative current
represent the interplay of two effects:
elastic scattering off impurities and the photovoltaic effect;
electric field in the domain is found from the condition that
 these two effects
compensate each other on average. However,
 those processes are statistically independent.
Consequently, this statistical independence of two
processes may be revealed through the
current noise in the ZRS or ZCS, which is not expected to
have any singularity in this regime. The analysis of this
noise can be performed by slight modification of the equations
derived in the present paper in the spirit of {\em e.g.} Ref.
~\onlinecite{Agam} and left as a subject for the future research.

\section*{Acknowledgements}
We are grateful to B.L. Altshuler and to A.V. Andreev
for participation in work \cite{memoryeffect} from which
essential part of the physics discussed in the present
paper was understood,
to V.I. Ryzhii for informing us about
Refs.~\onlinecite{Ryzhii,Ryzhii1,Zakharov,Ryzhii2}.
We thank V.I. Falko, A.J. Millis and M. Zudov
for reading the manuscript and valuable remarks.
Useful discussions with L.I. Glazman, A.I. Larkin and especially
with A.D. Mirlin are gratefully acknowledged.
I. Aleiner was supported by Packard foundation.
M. Vavilov was supported by NSF grants DMR01-20702, DMR02-37296,
and EIA02-10736.


\begin{thebibliography}{99}
\bibitem{Andoreview}See T. Ando, A.B. Fowler, and F. Stern,
  Rev. Mod. Phys, {\bf 54}, 437 (1982) for general review.

\bibitem{QHEreview}See {\em The Quantum Hall Effect},
R.E. Prange and S.M. Girvin (eds.), 2nd ed. (1990)
for general review.

\bibitem{Zudov1} M.A. Zudov, R.R. Du, J.A. Simmons, and J.L. Reno;
cond-mat/9711149; Phys. Rev. B {\bf 64}, 201311(R) (2001).

\bibitem{Mani} R. Mani, J.H. Smet, K. von Klitzing, V. Narayanamurti,
W.B. Johnson, and V. Umansky, Nature, {\bf 420}, 646 (2002).

\bibitem{Zudov2} M. A. Zudov, R. R. Du, L. N. Pfeiffer, and K. W. West,
Phys. Rev. Lett. {\bf 90}, 046807 (2003).

\bibitem{Mani2}
 R.G. Mani, J.H. Smet,
K. von Klitzing, V. Narayanamurti, W.B. Johnson, and V. Umansky,
cond-mat/0303034.

\bibitem{Zudov4}  C.L. Yang,
M.A. Zudov, T.A. Knuuttila, R.R. Du, L.N. Pfeiffer, and K.W. West,
cond-mat/0303472.

\bibitem{Dorozhkin} S.I. Dorozhkin, JETP Lett. {\bf 77}, 577 (2003)
[Pis'ma ZhETF {\bf 77}, 681 (2003)], cond-mat/0304604.

\bibitem{Ryzhii} V.I. Ryzhii, Fiz. Tverd. Tela, {\bf 11}, 2577
  (1969);
[Sov. Phys. Solid State, {\bf 11}, 2078, (1970)].

\bibitem{Ryzhii1} V.I. Ryzhii, R.A. Suris, and B.S. Shchamkhalova,
Fiz. Tekh. Poluprovodn. {\bf 20}, 2078, (1986)
[Sov. Phys. Semiconductors, {\bf 20}, 1289 (1986)].

\bibitem{Durst}
Adam C. Durst, Subir Sachdev, N. Read, and S.M. Girvin,
Phys. Rev. Lett. {\bf 91}, 086803 (2003).

\bibitem{Andreev}  A.V. Andreev, I.L. Aleiner, and A.J. Millis,
Phys. Rev. Lett. {\bf 91}, 056803 (2003).


\bibitem{Gunn}For review, see {\em e.g.} E. Sch\"oll.
{\em Nonlinear Spatio-Temporal Dynamics and Chaos in Semiconductors},
(Cambrdige University Press, 2001).


\bibitem{Zakharov}A.L. Zakharov, Zh. Eksp. Teor. Fiz. {\bf 38}, 665 (1960)
[Sov. Phys. JETP, {\bf 11}, 478 (1960)].

\bibitem{Dyakonov} M.I. D'yakonov, Pis'ma Zh. Eksp. Teor. Fiz. {\bf
    39}, 158 (1984) [JETP Lett., {\bf 39}, 185 (1984)];
M.I. D'yakonov and A.S. Furman, Zh. Eksp. Teor. Fiz. {\bf
    87}, 2063 (1984) [Sov. Phys. JETP, {\bf 60}, 1191 (1984)].

\bibitem{Ruby}  P.I. Liao, A.M. Glass, and L.M. Humphrey, Phys. Rev.
B, {\bf 22}, 2276 (1980); S.A. Basun, A.A. Kaplyanskii, and
  S.P. Feofilov,  Pis'ma Zh. Eksp. Teor. Fiz. {\bf
    37}, 492 (1983) [JETP Lett., {\bf 37}, 586 (1983)].


\bibitem{Anderson} P.W. Anderson and W.F. Brinkman, cond-mat/0302129.

\bibitem{HeandShe} Junren Shi and X.C. Xie,
Phys. Rev. Lett. {\bf 91}, 086801 (2003).

\bibitem{Volkov} F.S. Bergeret, B. Huckestein, A.F. Volkov,
Phys. Rev. B {\bf 67}, 241303 (2003).

\bibitem{Mikhailov}S.A. Mikhailov, cond-mat/0303130; Since this paper
lacks any real analysis of the kinetic of electrons
near the edge, we do not classify it  as a sound theoretical
prediction,
and can not discuss the relation to our results.
Simplest counterexample is the parabolic confinement
potential when the assertions of cond-mat/0303130 are not consistent
with the Kohn theorem.

\bibitem{newRaikh} A. A. Koulakov and M. E. Raikh, cond-mat/0302465.

\bibitem{Rivera}P. H. Rivera and P. A. Schulz, cond-mat/0305019.

\bibitem{Mirlin-ac} I.A. Dmitriev, A.D. Mirlin, and D.G. Polyakov,
cond-mat/0304529.

\bibitem{inelastic} I.A. Dmitriev, M.G.~Vavilov, I.L. Aleiner,
  A.D.~Mirlin, and D.G.~Polyakov, cond-mat/0310668.


\bibitem{Baskin79} E.M. Baskin, L.I. Magarill, and M.V. Entin,
Zh. Exsp. Teor. Fiz, {\bf 75}, 723 (1978).
[Sov. Phys. JETP, {\bf 48}, 365 (1978)].

\bibitem{Belinicher} For a review, see
 V.I. Belinicher and B.I. Sturman,
Usp. Fiz. Nauk 130, 415 (1980) [Sov. Phys. Usp. 23, 199 (1980)].

\bibitem{Ando}T. Ando and Y. Uemura,  J. Phys. Soc. Jpn. {\bf 36}, 959 (1974).




\bibitem{Raikh} M.E. Raikh and T.V. Shahbazyan,
Phys. Rev. B {\bf 47}, 1522 (1993).


\bibitem{Ryzhii2}V.I. Ryzhii, R.A. Suris, and B.S. Shchamkhalova,
Fiz. Tekh. Poluprovodn. {\bf 20}, 1404, (1986)
[Sov. Phys. Semiconductors, {\bf 20}, 883 (1986)].

\bibitem{mirlin1}
A. D. Mirlin, J. Wilke, F. Evers, D. G. Polyakov, and P. W\"olfle
Phys. Rev. Lett. {\bf 83}, 2801 (1999).



\bibitem{memoryeffect}I.L. Aleiner, B.L. Altshuler,
and A.V. Andreev (unpublished).

\bibitem{Keldysh} L.V. Keldysh, Zh. Eksp. Teor. Fiz. {\bf 47}, 1945 (1964)
[Sov. Phys. JETP {\bf 20}, 1018 (1964)];
we use the Green function parametization of Ref.~\onlinecite{LO77}.

\bibitem{LO77} A.I. Larkin and Y.N. Ovchinnikov,
Zh. Eksp. Teor. Fiz., {\bf 73}, 299 (1977)
[Sov. Phys. JETP, {\bf 46}, 155 (1977)].

\bibitem{Eilenberger}G. Eilenberger, Z. Phys. Bd. {\bf 214}, 195
(1968).

\bibitem{68} A.I. Larkin and Y.N. Ovchinnikov,
Zh. Eksp. Teor. Fiz., {\bf 55}, 2262 (1968)
[Sov. Phys. JETP, {\bf 28}, 1200 (1969)].

\bibitem{Wegner79} F. Wegner, Z. Phys. B, {\bf 35}, 207 (1979).

\bibitem{Benedict} P. Carra, J.T. Chalker, K.A. Benedict,
Ann. Phys. NY, {\bf 194}, 1 (1989).



\bibitem{Altshuler}B.L. Altshuler, A.G. Aronov, and
  D.E. Khmelnitskii,
Solid State Commun. {\bf 39}, 619 (1981).

\bibitem{mathbook} I.S. Gradshteyn and I.M. Ryzhik.
{\em Tables of Integrals, Series, and Products}, 5th ed,
(Academic Press, 1994).


\bibitem{Ando1} T. Ando, J. Phys. Soc. Jpn. {\bf 37} 1233, (1974);
{\bf 37}, 622 (1974); {\bf 36}, 1521 (1974).






\bibitem{EAltshuler}B. Laikhtman and E.L. Altshuler,
Ann. Phys. NY,  {\bf 232}, 332 (1994).


\bibitem{Efetov} K.B. Efetov and V.G. Marikhin, Phys. Rev. B, {\bf
    40}, 12126 (1989).


\bibitem{Agam} O. Agam, I.L. Aleiner, and A.I. Larkin,
Phys. Rev. Lett, {\bf 85}, 3153 (2000).



\end{thebibliography}
\end{document}